\documentclass[11pt,a4paper]{article}
\pdfoutput=1
\usepackage{jheppub}
\usepackage{rotating}
\usepackage{array}
\usepackage{amsmath}
\usepackage{slashed,hyperref}
\usepackage[normalem]{ulem}

\newcommand{\alp}{a}
\newcommand{\ga}{g_{a\gamma}}

\preprint{CERN-PH-TH-2015-293, DESY 15-237}

\title{ALPtraum: ALP production in proton beam dump experiments}

\author[a]{Babette D\"obrich,}
\author[b]{Joerg Jaeckel,}
\author[c]{Felix Kahlhoefer,}
\author[c]{Andreas Ringwald,}
\author[c]{and \mbox{Kai Schmidt-Hoberg}}

\affiliation[a]{CERN, 1211 Geneva 23, Switzerland}
\affiliation[b]{Institut f\"{u}r Theoretische Physik, Universit\"at Heidelberg, Philosophenweg 16, 69120 Heidelberg, Germany}
\affiliation[c]{DESY, Notkestrasse 85, 22607 Hamburg, Germany}

\emailAdd{babette.dobrich@cern.ch}
\emailAdd{jjaeckel@thphys.uni-heidelberg.de}
\emailAdd{felix.kahlhoefer@desy.de}
\emailAdd{andreas.ringwald@desy.de}
\emailAdd{kai.schmidt.hoberg@desy.de}

\abstract{With their high beam energy and intensity, existing and near-future proton beam dumps provide an excellent opportunity to search for new very weakly coupled particles in the MeV to GeV mass range. One particularly interesting example is a  so-called axion-like particle (ALP), i.e.\ a pseudoscalar coupled to two photons. The challenge in proton beam dumps is to reliably calculate the production of the new particles from the interactions of two composite objects, the proton and the target atoms. In this work we argue that Primakoff production of ALPs proceeds in a momentum range where production rates and angular distributions can be determined to sufficient precision using simple electromagnetic form factors. Reanalysing past proton beam dump experiments for this production channel, we derive novel constraints on the parameter space for ALPs. We show that the NA62 experiment at CERN could probe unexplored parameter space by running in `dump mode' for a few days and discuss opportunities for future experiments such as SHiP.}


\begin{document}

\maketitle
\flushbottom

\section{Introduction}

Over the last few years Nature has been quite reluctant to reveal its secrets beyond the Standard Model~--- certainly not for a lack of trying on our side. While the second run of the LHC still promises significant discovery potential, it also seems sensible to carefully examine our search strategy and look for potential places we have missed as well as for new opportunities that can be explored in parallel. In this endeavour it is only prudent that we first look at technologies and apparatuses that are already at our disposal and then turn to more ambitious future projects. 

In this note we want to discuss an explicit and particularly promising example of this strategy: we will show that without major modifications an existing proton fixed target experiment, 
NA62~\cite{Hahn:1404985} at the CERN SPS, offers discovery potential for axion-like particles (ALPs) in the MeV to GeV range and can be competitive with existing experiments in a rather short run-time. A proposed dedicated search at the SPS, SHiP~\cite{Anelli:2015pba,Alekhin:2015byh}, could then probe even deeper into untested parameter space.

Fixed target experiments are particularly useful to search for new weakly coupled particles in the MeV to GeV range~\cite{Bjorken:1988as,Blumlein:1990ay,Bjorken:2009mm,Andreas:2012mt}, because they nicely combine a sufficiently energetic reaction to produce particles in this mass regime with a sufficiently high reaction rate to probe small couplings. The two most common types of beams are electron and proton beams. Indeed some of the best current limits on axion-like particles in the MeV to GeV mass range originate from electron beam fixed-target experiments~\cite{Konaka:1986cb,Riordan:1987aw,Bjorken:1988as,Davier:1989wz} as well as electron-positron colliders~\cite{Jaeckel:2015jla}. As demonstrated in~\cite{Blumlein:1990ay,Bergsma:1985qz, Batell:2009di}, another promising option are proton fixed-target experiments, in particular those making use of the high intensity of 400 GeV protons from the SPS at CERN. It is therefore a worthwhile task to determine the sensitivity of existing experiments, such as the NA62 experiment, as well as proposed experiments especially optimised for the search of long-lived neutral particles, such as SHiP.

To fully exploit these experiments for the search of axion-like particles one needs a reliable calculation of the production rates and angular distributions. Due to the composite nature of both the proton and the nucleus this seems particularly challenging. However, we argue that for a high-energy proton beam one can reliably calculate the cross section for the fusion of two photons into one ALP (so-called Primakoff production~\cite{Halprin:1966zz}) for ALPs in the interesting MeV to GeV mass range using simple atomic form factors. The reason for this is as follows: Both the proton and the nucleus are surrounded by the virtual photons that make up the usual electric field of a charged particle. Due to the highly relativistic nature of the incoming protons, the photons from the proton now `see' the photons from the nucleus as highly blue shifted, thereby being able to provide more energy/momentum. This effect allows to create relatively massive ALPs from photons that, in the rest frame of the proton/nucleus, are too soft to be affected by the sub-structure of the proton/nucleus. In other words, the momentum transfer in the respective rest frames is sufficiently small that we can use simple electromagnetic form factors to describe the photon interactions.

At the same time, for the masses of interest to us, the typical photons from the nucleus have sufficient momentum to be essentially unaffected by the shielding from the electron shell, leading to an enhancement of the cross section proportional to the square of the nucleus charge. Coherent Primakoff production in this Goldilocks zone can therefore dominate over incoherent production processes and lead to detectable signals in present and near-future experiments for a significant range of ALP parameter space that is not yet explored. Crucially, it is possible for us to reliably calculate not only the production rate, but also angular distributions, which are particularly important to obtain a realistic calculation of the geometric acceptances that determine the sensitivity of a real experiment. 

This paper is structured as follows. In section~\ref{sec:review} we review the nature of and motivation for ALPs and for searches in the low-energy/high-intensity regime. In section~\ref{sec:production} we then provide a detailed calculation of the Primakoff production of ALPs from high energy protons impinging on a fixed target, paying particular attention to the angular distributions. In section~\ref{sec:eventrates} we proceed to discuss the general features of the production and decay relevant for the calculation of the experimental acceptances before we use them in section~\ref{sec:constraints} in order to derive new constraints on the ALP parameter space from CHARM~\cite{Bergsma:1985qz} and NuCal~\cite{Blumlein:1990ay}. In section~\ref{sec:na62} we then estimate the sensitivity of a potential run at NA62, followed by a similar analysis for the proposed SHiP experiment in section~\ref{sec:SHiP}. We conclude in section~\ref{sec:conclusions}.

\section{A brief review and motivation of ALPs \label{sec:review}}

Progress in particle physics has been guided by the paradigm of renormalizable interactions with $\mathcal{O}(1)$ dimensionless couplings. This paradigm suggests that any new particle to be discovered should be heavy and reveal its presence either directly in high-energy colliders or indirectly by mediating higher-dimensional interactions, which could then be probed in precision measurements such as muon $g-2$ experiments or flavour experiments. It has however become increasingly clear that there are other options. Even light particles could still remain to be discovered, provided they have sufficiently small interactions with Standard Model (SM) particles, and therefore with our experiments.

Popular examples for such Pseudo-Goldstone bosons are axion-like particles, which are loosely defined as (pseudo-)scalar particles coupled to the SM particles by dimension-5 couplings to two gauge bosons or derivative interactions to fermions (the so-called axion portal~\cite{Nomura:2008ru,Batell:2009di}). Light pseudoscalars have received a considerable amount of interest recently in the context of dark matter model-building, because they may act as a mediator for the interactions between dark matter and SM particles. In these models it is easily possible to reproduce the observed dark matter relic abundance from thermal freeze-out while evading the strong constraints from direct and indirect detection experiments~\cite{Boehm:2014hva,Berlin:2015wwa}.

In the present work we focus on pseudoscalar ALPs  whose dominant interaction is with photons\footnote{Other possibilities are discussed in section~\ref{sec:conclusions}.} and we are interested in masses and energies of the order of MeV to GeV, significantly below the scale of electroweak symmetry breaking. The relevant Lagrangian is then
\begin{equation}
\mathcal{L}= \frac{1}{2} \partial^\mu a \, \partial_\mu a - \frac{1}{2}m^{2}_{\alp} \, \alp^2-\frac{1}{4} \, \ga \, \alp \, F^{\mu\nu}\tilde{F}_{\mu\nu} \; ,
\end{equation}  	
where $\ga$ denotes the photon ALP coupling.

The origin of such an effective coupling to photons can be motivated in analogy to the case of the axion and is generic for any (pseudo-)Goldstone boson of an axial symmetry with non-vanishing electromagnetic anomaly. One naturally expects the effective coupling to be of order
\begin{equation}
\ga \sim \frac{\alpha}{2\pi F} \; .
\end{equation}
Probing a small value of $\ga$ therefore effectively allows us to probe a large scale $F$ of the underlying more fundamental theory from which the pseudo-Goldstone boson arises. We emphasise that we take the ALP mass and the ALP-photon couplings to be independent parameters.

In terms of the effective ALP-photon coupling $\ga$, the ALP decay width is\footnote{We assume that the total decay width is dominated by the ALP-photon coupling $\ga$ and that couplings to leptons and quarks~--- if present~--- give subleading contributions. We discuss the impact of additional decay channels in section~\ref{sec:conclusions}.}
\begin{equation}
\Gamma = \frac{\ga^2 \, m_{a}^3}{64 \pi}
\end{equation}
and the ALP lifetime is given by $\tau = 1/\Gamma$. For an ALP with energy $E_a \gg m_a$ in the laboratory frame, the typical decay length is then given by
\begin{align}
\label{decaylength}
l_a & = \beta \, \gamma \, \tau \approx \frac{64 \pi \, E_a}{\ga^2 \, m_{a}^4} \nonumber \\
& \approx 40\:\text{m} \times \frac{E_a}{10\:\text{GeV}} \left(\frac{\ga}{10^{-5}\:\text{GeV}^{-1}}\right)^{-2} \left(\frac{m_a}{100\:\text{MeV}}\right)^{-4} \; .
\end{align}

\begin{figure}[tb]
\centering
\includegraphics[width=0.47\textwidth]{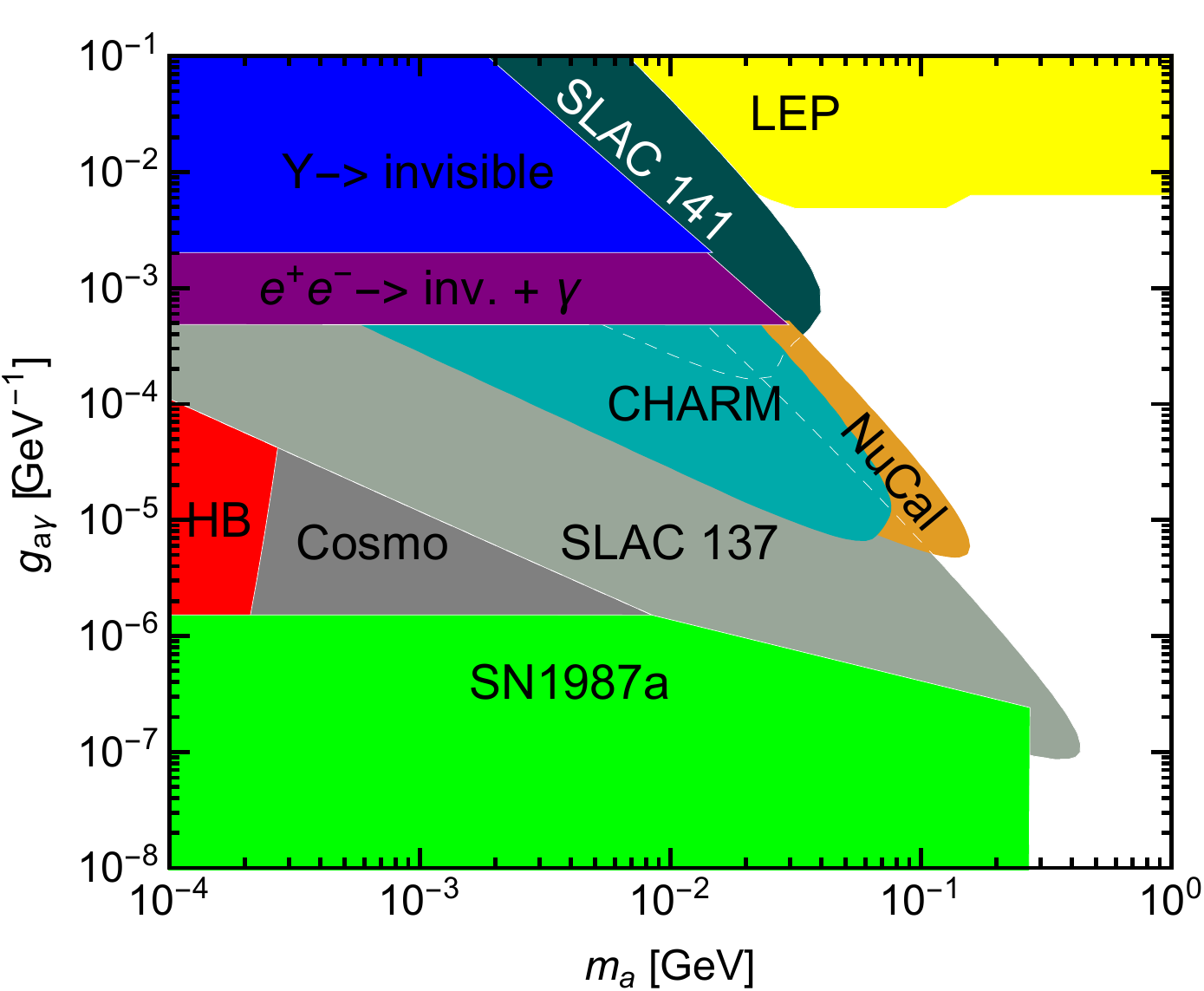}
\caption{
Summary of constraints on the ALP parameter space (compilation from~\cite{Jaeckel:2015jla} and references therein; in particular SLAC electron fixed target limits are from~\cite{Riordan:1987aw,Bjorken:1988as,Hewett:2012ns}). The new limits from the proton beam dump experiments CHARM and NuCal, derived in the present paper, are shown in turquoise and orange.
}
\label{fig:exclusions}
\end{figure}

A given experiment will be most sensitive to ALPs with a decay length comparable to the distance $L$ between target and the detector. Particles with shorter decay length are likely to decay before they reach the decay volume and the decay products will be absorbed. Crucially, larger couplings imply shorter decay lengths and therefore lead to an exponential suppression of the expected number of events in a given experiment. 
It is therefore a great challenge to probe ALP-photon couplings in the range $10^{-6}\:\text{GeV}^{-1} < \ga < 10^{-2}\:\text{GeV}^{-1}$ for ALP masses above $10\:\text{MeV}$ (cf.\ figure~\ref{fig:exclusions}).\footnote{Both smaller couplings and smaller ALP masses are in fact very strongly constrained by astrophysical and cosmological observations. Larger couplings, on the other hand, can be tested directly at colliders such as LEP or the LHC~\cite{Jaeckel:2015jla}.} While these couplings are large enough to produce a significant number of ALPs in the target of a beam dump experiment, the fraction of ALPs that reach the detector depends sensitively on the detector geometry and the beam energy. The higher the beam energy and the shorter the distance between target and detector, 
the larger ALP-photon couplings can be probed. The high beam energy of proton beam dump experiments is therefore suited for making progress in the large coupling window. This effect can be seen in figure~\ref{fig:exclusions} (cf.\ section~\ref{sec:constraints} for details). The turquoise region from the proton beam dump experiment CHARM extends beyond the limit from the electron beam dump experiment SLAC 137, even though the former has a longer distance to the decay volume. Nevertheless, this can only partially compensate the geometric limitations. For better sensitivity an improved geometry as present in NA62 and even more so at SHiP is essential.

For small couplings and masses the decay length is longer than the
{distance from the ALP production point to the most down-stream detectors in the experiment}. Hence only a fraction of ALPs decay such that the resulting photons can be detected in the detector (cf.\ eq.~\eqref{decaylength}). This leads to the typical diagonal ellipse for the probed regions as shown in figure~\ref{fig:exclusions}.

\section{ALP production in proton beam dumps \label{sec:production}}

\begin{figure}[tb]
\centering
\includegraphics[width=0.4\textwidth,clip,trim=150 570 300 140]{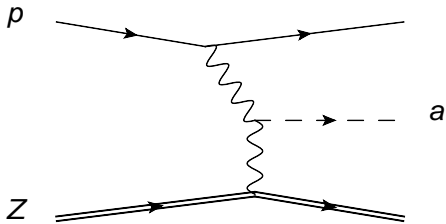}
\caption{Primakoff production of ALPs in proton-nucleus collisions.}
\label{fig:Primakoff}
\end{figure}

Proton beam dumps can produce ALPs via coherent scattering (i.e.\ elastic scattering of the entire proton on the nucleus in the target), incoherent scattering (i.e.\ scattering of individual quarks or gluons from the proton and/or the nucleus) and non-perturbative processes (e.g.\ in the decay of hadronic resonances). The best example for coherent scattering is the Primakoff process shown in figure~\ref{fig:Primakoff}. As we will see, in contrast to incoherent scattering and non-perturbative processes, the Primakoff process remains perturbative at low energies and is not affected by hadronic uncertainties, so that predictions can be calculated with good accuracy. Moreover, the Primakoff process is particularly interesting for beam dump experiments for the following reasons:
\begin{enumerate}
 \item The cross section for coherent scattering is typically proportional to the square of the charge of the 
 target nucleus, $Z^2$, while incoherent scattering and non-perturbative processes are roughly proportional 
 to the mass number of the target nucleus, $A$. For heavy target materials, such as lead or tungsten,
 this can lead to a very significant enhancement of the rate of ALP production. But even for medium-weight target materials like copper, iron or molybdenum Primakoff production is still competitive with other production mechanisms~\cite{Blumlein:1990ay}.
 \item The typical momentum transfer in coherent scattering is small, such that the ALPs produced via the Primakoff process typically have very small transverse momenta (i.e.\ momenta perpendicular to the direction of the beam). As a result, cross sections are very strongly peaked in the forward direction, such that even a relatively small detector far away from the target can have a large geometric acceptance. This makes it possible to use the Primakoff process to search for ALPs with very large decay length.
\end{enumerate}
Primakoff production is therefore not only the theoretically cleanest production mode, but can be expected to give the dominant contribution for ALPs that do not directly couple to quarks.\footnote{For ALPs with direct quark couplings stronger constraints are expected to arise from flavour-changing rare decays, such as $K \rightarrow \pi + a$~\cite{Hiller:2004ii,Freytsis:2009ct,Andreas:2010ms,Dolan:2014ska}. However, it is very difficult to accurately predict the rate and distribution of ALPs produced in this way.} In the present work we will therefore focus on this production mechanism. 

The Primakoff process can be studied in two different ways. In the laboratory frame, where the target nucleus is at rest, it makes sense to calculate the equivalent photon spectrum for the proton beam and then consider the process $\gamma + N \rightarrow a + N$, i.e.\ calculate the probability for the photon to emit an ALP before being absorbed by the nucleus. In the centre-of-mass frame, on the other hand, both the proton and the target nucleus are moving, so one can calculate the equivalent photon spectrum for each and then consider the photon-fusion process $\gamma + \gamma \rightarrow a$. The former approach was considered previously in~\cite{Bjorken:1988as} for the analysis of electron beam dumps, the latter approach was employed by~\cite{Blumlein:1990ay} for a proton beam dump experiment.

Of course, both approaches are physically equivalent and should lead to identical predictions for the ALP production cross section and distribution. However, as we will discuss in detail below, in order to accurately predict the angular distribution of the ALPs produced in the beam dump, it is essential to properly take into account the transverse momentum of the virtual photons (see e.g.~\cite{Blumlein:2011mv,Blumlein:2013cua}). These transverse momenta have been neglected entirely in~\cite{Blumlein:1990ay}, whereas in~\cite{Bjorken:1988as} only the virtual photon connected to the target nucleus is allowed to have non-zero transverse momentum.\footnote{As we will see below, this is a valid approximation for electron beam dump experiments, because the transverse momentum of the virtual photon cannot significantly exceed the mass of the radiating particle. However, the approximation is not sufficient when considering the case of proton beam dump experiments.} We therefore revisit both approaches, carefully taking into account all transverse momenta, and derive the angular distribution for ALPs produced in proton beam dumps.

\subsection{The equivalent photon approximation \label{sec:epa}}

The equivalent photon approximation (EPA), also known as the Weizs\"{a}cker-Williams approximation, provides a convenient framework for studying processes involving photons emitted from fast-moving charges. The basic idea is to replace the charged particle(s) in the initial state by photons following a distribution $\gamma(\omega, q_t)$ and then consider directly the interactions of these photons. Since the transverse momentum $q_t$ of the photons is typically very small, it is common in the literature to integrate the distribution $\gamma(\omega, q_t)$ over $q_t$ and consider only the energy distribution $\gamma(\omega)$, assuming that all photons travel in the same direction as the charged particle they originated from.\footnote{This is for example the case for the photon-from-proton mode 
implemented in MadGraph 5, version 2.3.3~\cite{Alwall:2011uj}.} For our purposes, however, it will be crucial to accurately predict angular distribution down to very small scattering angles, so that we need to include the distribution of transverse momenta. Our discussion follows ref.~\cite{Budnev:1974de} (see ref.~\cite{Martin:2014nqa} for a recent review).

Let us consider a particle of mass $m$ and energy $E$ and define the energy fraction of the photon as $x \equiv \omega / E$. The magnitude of the momentum transfer between particle and photon is given by $q^2 = |t|$, where
\begin{equation}
t(x, q_t) = -\frac{q_t^2 + x^2 \, m^2}{1-x} \; .
\end{equation}
Defining $q^2_\text{min} = -t(x, 0)$, the distribution of photons can then be written as
\begin{align}
\label{eq:epa}
 \gamma(x, q_t^2) \equiv \frac{\mathrm{d}^2 n_\gamma(x, q_t)}{\mathrm{d}x \, \mathrm{d}q_t^2} & = \frac{\alpha}{\pi} \frac{1}{x(1-x)} \frac{1}{q^2}\left[1-x+\frac{x^2}{2}-(1-x)\frac{q^2_\text{min}}{q^2} \right] \nonumber \\
& = \frac{\alpha}{2\pi} \frac{1+(1-x)^2}{x} \frac{q_t^2}{(q_t^2+x^2\,m^2)^2} \; ,
\end{align}
where $\alpha$ is the fine-structure constant and we have dropped a term proportional to $x^3$ in the second line, assuming $x \ll 1$. We note in particular that transverse momenta are predicted to be of order $q_t \sim x \, m \ll m$. This observation makes clear why it is a good approximation to neglect transverse momenta for the equivalent photon spectra of electron beams.

The equation above is valid for a point-like particle with unit charge. For protons or ions it needs to be modified to include the appropriate electromagnetic form factors:
\begin{align}
 \gamma(x, q_t^2) = \frac{\alpha}{2\pi} \frac{1+(1-x)^2}{x} \left[\frac{q_t^2}{(q_t^2+x^2\,m^2)^2} D(q^2) + \frac{x^2}{2} C(q^2)\right] \; ,
\end{align}
For protons, we follow~\cite{Budnev:1974de} and take
\begin{equation}
 D(q^2) = \frac{4 \, m_p^2 \, G_\text{E}^2(q^2) + q^2 \, G_\text{M}^2(q^2)}{4\,m_p^2 + q^2} \; , \quad C(q^2) = G_\text{M}^2(q^2) \; ,
\end{equation}
where
\begin{equation}
G_\text{E}(q^2) = \frac{1}{(1+q^2/q_0^2)^2} \; , \quad G_\text{M}(q^2) = \frac{\mu_p}{(1+q^2/q_0^2)^2}
\end{equation}
with $\mu_p^2 = 7.78$ the magnetic moment of the proton and $q_0^2 = 0.71\:\text{GeV}^2$. To make a conservative estimate we set the both form factors to zero for $q > 1\:\text{GeV}$. 

For heavy ions with charge number $Z$ and mass number $A$, one finds
\begin{equation}
 D(q^2) = Z^2 \, F(q^2)^2 \; , \quad C(q^2) \approx \mu_N^2 F_\text{M}(q^2)^2 \; ,
\end{equation}
where $F$ and $F_\text{M}$ are the charge and magnetic moment form factors, respectively. Since the contribution from the magnetic moment is not proportional to $Z^2$, it can be neglected to good approximation. For the charge form factor we use the Helm form factor, i.e.\ 
\begin{equation}
\label{eq:formfactor}
F(q^2) = \frac{3 \, j_1(\sqrt{q^2} R_1)}{\sqrt{q^2} R_1} \exp\left[-\frac{(\sqrt{q^2}\,s)^2}{2}\right] \; ,
\end{equation}
where $j_1$ is the first spherical Bessel function of the first kind and we employ a parametrization as provided in \cite{Lewin:1995rx},
with
\begin{equation}
R_1 = \sqrt{(1.23 \, A^{1/3}-0.6)^2+\frac{7}{3} \pi^2 \, 0.52^2-5 \, s^2} \; , 
\end{equation}
where $s=0.9\:\text{fm}$. The Helm form factor vanishes for $q \, R_1 = 4.49$ 
 and we set it to zero for $q > 4.49 / R_1$. 

For scattering on neutral atoms one furthermore needs to take into account screening of the nuclear 
charge by electrons. These screening effects will become important for small $q^2$, implying that one has 
to consider the electromagnetic form factor of the entire atom rather than the one of the nucleus. 
In principle, it is straight-forward to extract the atomic form factor at smaller momentum transfer 
from measurements (see e.g.~\cite{Henke:1993eda}). 
At energies above the K-shell threshold we do not expect large deviations from $D(q^2) = Z^2$, but even for smaller energies the deviations may be relatively small. Comparing to experimentally measured form factors in the literature~\cite{nist}, we conclude that for molybdenum it is a good approximation to take $D(q^2)=Z^2$ for $q>20$~keV, while for copper and iron $D(q^2)=Z^2$ holds for $q>10$~keV. For smaller momenta we set the form factor equal to zero, 
which has a negligible effect on the predicted cross section.

\subsection{Method 1: Photon fusion}

Let us first consider the production of ALPs in the centre-of-mass frame. If the proton has momentum $p_p$ in the laboratory frame and we denote the mass of the target nucleus with $m_N \gg m_p$, the centre-of-mass energy is approximately given by
\begin{equation}
 s \approx m_N^2 + 2 \, p_p \, m_N \; .
\end{equation}
The momentum of the two particles in the centre-of-mass frame is therefore
\begin{equation}
 p^\text{cms} \approx \sqrt{\frac{p_p^2 \, m_N}{m_N + 2 \, p_p}}
\end{equation}
and the corresponding energies are given by $E_i = \sqrt{(p^\text{cms})^2 + m_i^2}$. Denoting the respective equivalent photon spectra by $\gamma_p(x_1, q_{t,1}^2)$ and $\gamma_N(x_2, q_{t,2}^2)$, where $x_1 = \omega_1 / E^\text{cms}_p$ and $x_2 = \omega_2 / E^\text{cms}_N$, the total cross section for ALP production is then given by
\begin{equation}
 \sigma_{pN} = \int \mathrm{d}x_1 \, \mathrm{d}x_2 \, \mathrm{d}q_{t,1}^2 \, \mathrm{d}q_{t,2}^2 \, \gamma_p(q_{t,1}^2,x_1) \, \gamma_N(q_{t,2}^2,x_2) \, \sigma(\gamma \gamma \rightarrow a) \; ,
\end{equation}
where the cross section for the process $\gamma + \gamma \rightarrow a$ is calculated in appendix~\ref{app:crosssections} to be
\begin{equation}
 \sigma(\gamma \gamma \rightarrow a) = \frac{\pi\,\ga^2\,m_a}{16} \delta(m_{\gamma\gamma} - m_a) \; .
\end{equation}

The invariant mass of the two-photon system is given by
\begin{equation}
m_{\gamma\gamma} = 2 \sqrt{x_1\,x_2} \sqrt{E^\text{cms}_p \, E^\text{cms}_N} \; . 
\end{equation}
Moreover, the momentum of the ALP along the direction of the beam is given by $k_z = x_1 \, E^\text{cms}_p - x_2 \, E^\text{cms}_N$. Changing variables from $x_1$ and $x_2$ to $m_{\gamma\gamma}$ and $k_z$ and performing the integration over $m_{\gamma\gamma}$ we obtain 
\begin{align}
 \sigma_{pN} = & \frac{\pi\,\ga^2\,m_a}{16} \int \mathrm{d}k_z \, \mathrm{d}q_{t,1}^2 \, \mathrm{d}q_{t,2}^2 \, \frac{m_a}{2 \, E^\text{cms}_p \, E^\text{cms}_N \sqrt{k_z^2 + m_a^2}} \nonumber \\ & \times \gamma_p\Bigl( q_{t,1}^2, \frac{\sqrt{k_z^2 + m_a^2} + k_z}{2\,E^\text{cms}_p} \Bigr) \, \gamma_N \Bigl(q_{t,2}^2, \frac{\sqrt{k_z^2 + m_a^2} + k_z}{2\,E^\text{cms}_N}\Bigr)  \; .
\end{align}

\begin{figure}[tb]
\centering
\includegraphics[width=0.47\textwidth]{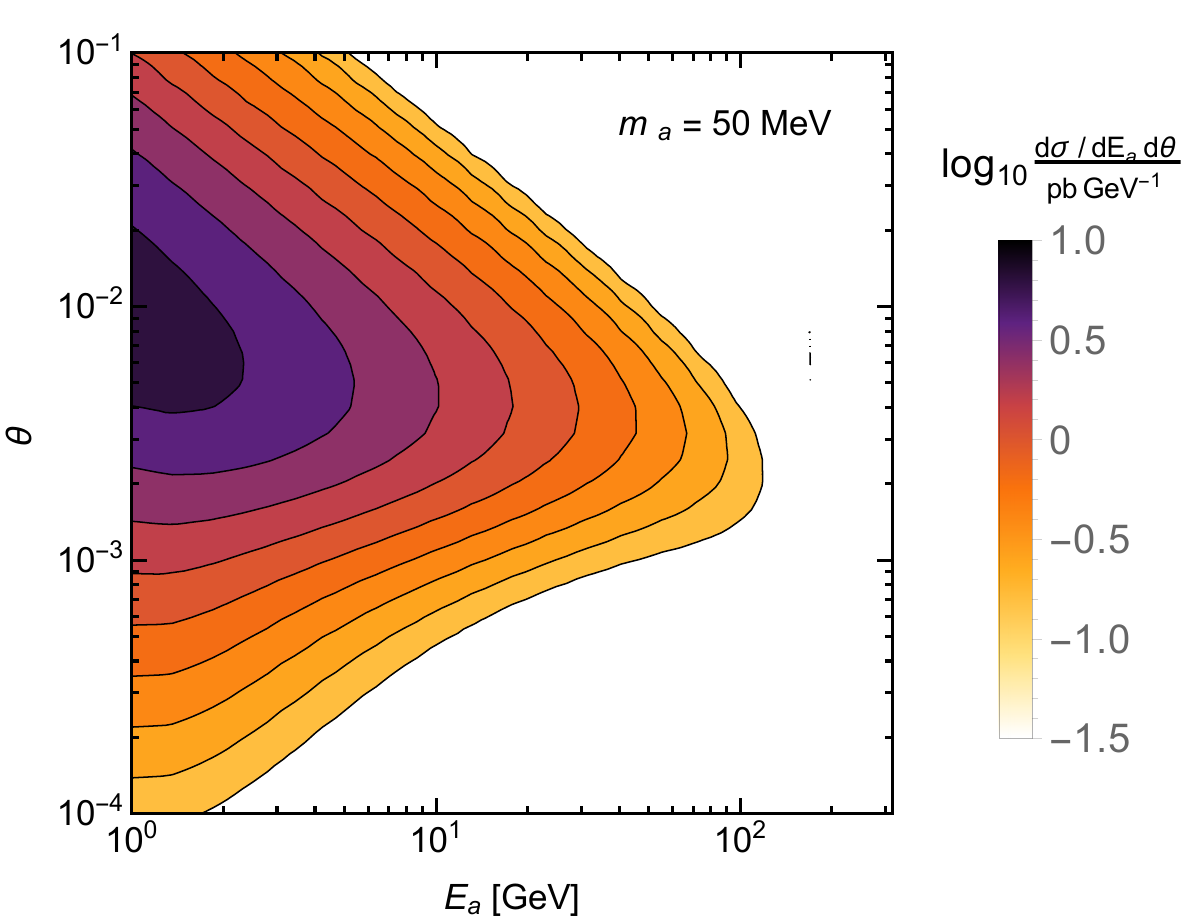}\hfill\includegraphics[width=0.47\textwidth]{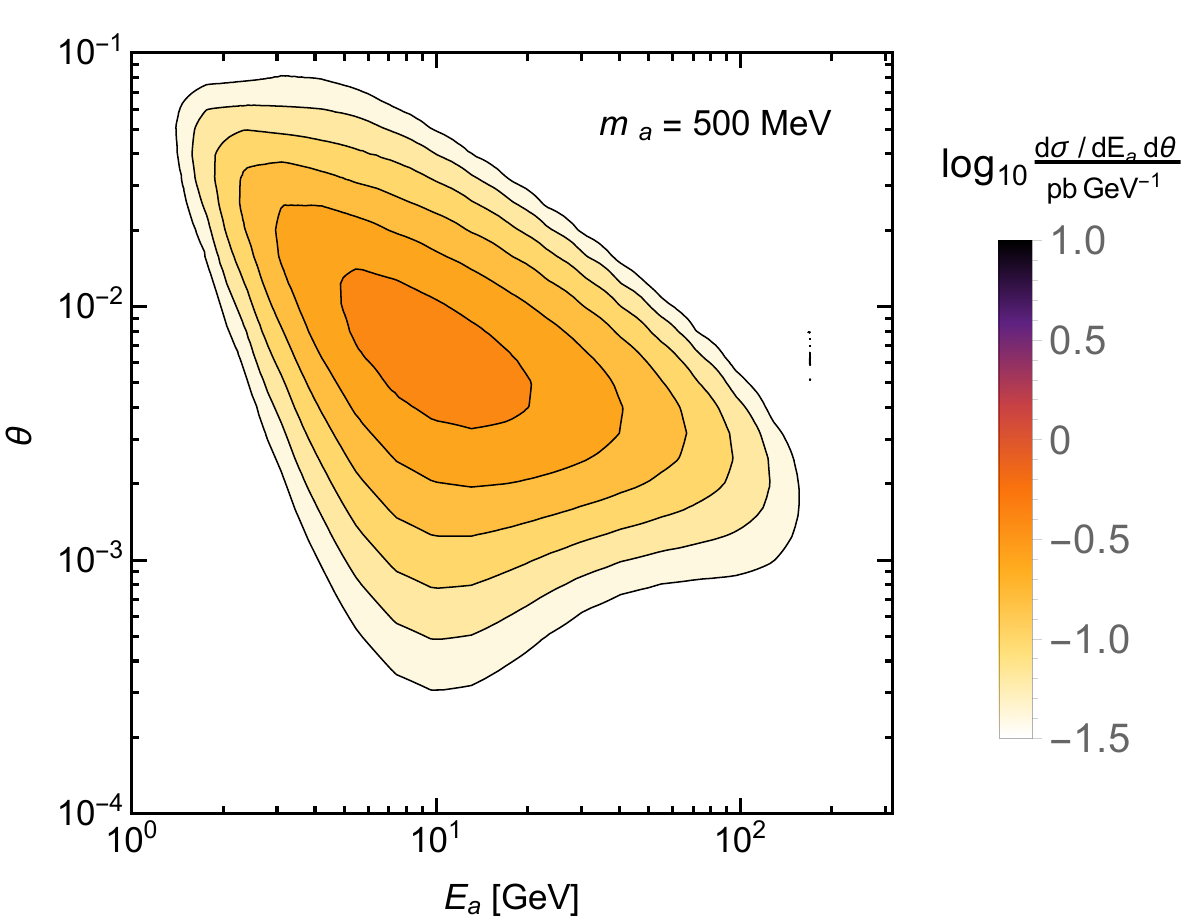}
\caption{Predictions for the differential ALP-production cross section from 400~GeV
protons on a copper target in the laboratory frame at $g_{a \gamma}= 10^{-4}\ {\rm GeV}^{-1}$ for 
$m_a = 50\:\text{MeV}$ (left) and $m_a = 500\:\text{MeV}$ (right) .}
\label{fig:alp-distribution}
\end{figure}

It is straight-forward from this expression to extract the momentum distribution of ALPs along the direction of the beam, $\mathrm{d}\sigma_{pN}/\mathrm{d}k_z$. However, as discussed above, we are also interested in the momentum distribution perpendicular to the direction of the beam, $k_t$. Calculating this distribution is complicated by the fact that the transverse momenta of the two photons do not have to be parallel and therefore need to be added vectorially $\mathbf{k}_t = \mathbf{q}_{t,1} + \mathbf{q}_{t,2}$. Nevertheless, assuming that the angle between the two photons is uniformly distributed between 0 and $2\pi$, it is possible to analytically average over the photon-photon angle. As shown in appendix~\ref{app:angular}, we obtain
\begin{align}
\frac{\mathrm{d}^2\sigma_{pN}}{\mathrm{d}k_z \, \mathrm{d}k_t^2} = & \frac{\ga^2\,m_a^2}{32 \, E^\text{cms}_p \, E^\text{cms}_N \sqrt{k_z^2 + m_a^2}}  \nonumber \\ & \times \int_A \mathrm{d}q_{t,1}^2 \mathrm{d}q_{t,2}^2 \frac{\gamma_p \Bigl(q_{t,1}^2, \frac{\sqrt{k_z^2 + m_a^2} + k_z}{2\,E^\text{cms}_p} \Bigr) \, \gamma_N\Bigl(q_{t,2}^2, \frac{\sqrt{k_z^2 + m_a^2} + k_z}{2\,E^\text{cms}_N}\Bigr)}{\sqrt{2 \, k_t^2 \, (q_{t,1}^2 + q_{t,2}^2) - k_t^4 - (q_{t,1}^2-q_{t,2}^2)^2}} \; \; ,
\label{eq:method1}
\end{align}
where the area of integration $A$ is defined by $|q_{t,1} - q_{t,2}| \leq k_t \leq q_{t,1} + q_{t,2}$. The integral can easily be evaluated numerically. The resulting distribution can then be transformed into the laboratory frame, in order to determine $\mathrm{d}^2 \sigma_{pN} / \mathrm{d} E_a \, \mathrm{d} \cos \theta$. We show two examples for the resulting distributions for different ALP masses in figure~\ref{fig:alp-distribution} at $g_{a \gamma}= 10^{-4}\ {\rm GeV}^{-1}$.

\subsection{Method 2: Photon absorption}

An alternative way to calculate the rate and distribution of ALPs produced in a proton beam dump experiment is to consider the equivalent photon spectrum of the proton beam in the laboratory frame and calculate the probability that a photon emits an ALP before being absorbed by the nucleus. In other words, we need to calculate the cross section for the $2\rightarrow2$ process $\gamma + N \rightarrow a + N$ and multiply with the distribution of photon momenta $\gamma_p(x_1, q_t^2)$, where now $x_1 = \omega_1 / E_\text{beam}$. As shown in appendix~\ref{app:crosssections}, the relevant cross section is approximately given by
\begin{align}
\frac{\mathrm{d}\sigma_{\gamma N}}{\mathrm{d}\cos{\theta}} & \simeq \frac{\alpha \, \ga^2 \, (- 4 \, E_a^2 \, t - m_a^4)}{16 \, t^2} Z^2 \, F(|t|)^2\; ,
\end{align}
where $F(|t|)^2$ is the form factor for the target nucleus (see eq.~(\ref{eq:formfactor})) and
\begin{equation}
 t = -\frac{m_a^4}{4 \, E_a^2} - p_t^2 + 2 \, E_a \, p_t \, \theta \, \cos \phi - E_a^2 \, \theta^2 \; ,
\end{equation}
with $p_t$ the transverse photon momentum, $\theta$ the angle between the ALP momentum and the beam direction and $\phi$ the angle between the transverse momenta of ALP and photon.

\begin{figure}[tb]
\centering
\includegraphics[width=0.47\textwidth]{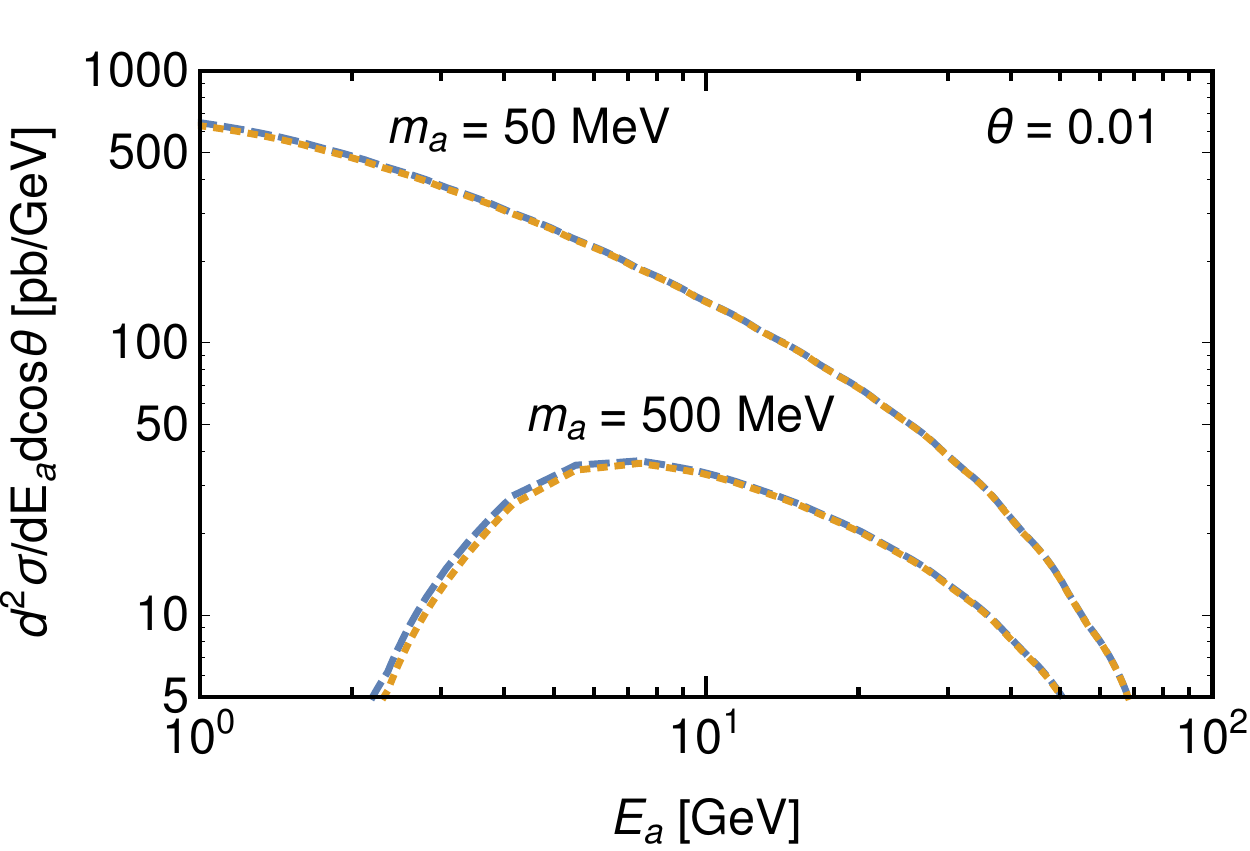}\hfill\includegraphics[width=0.47\textwidth]{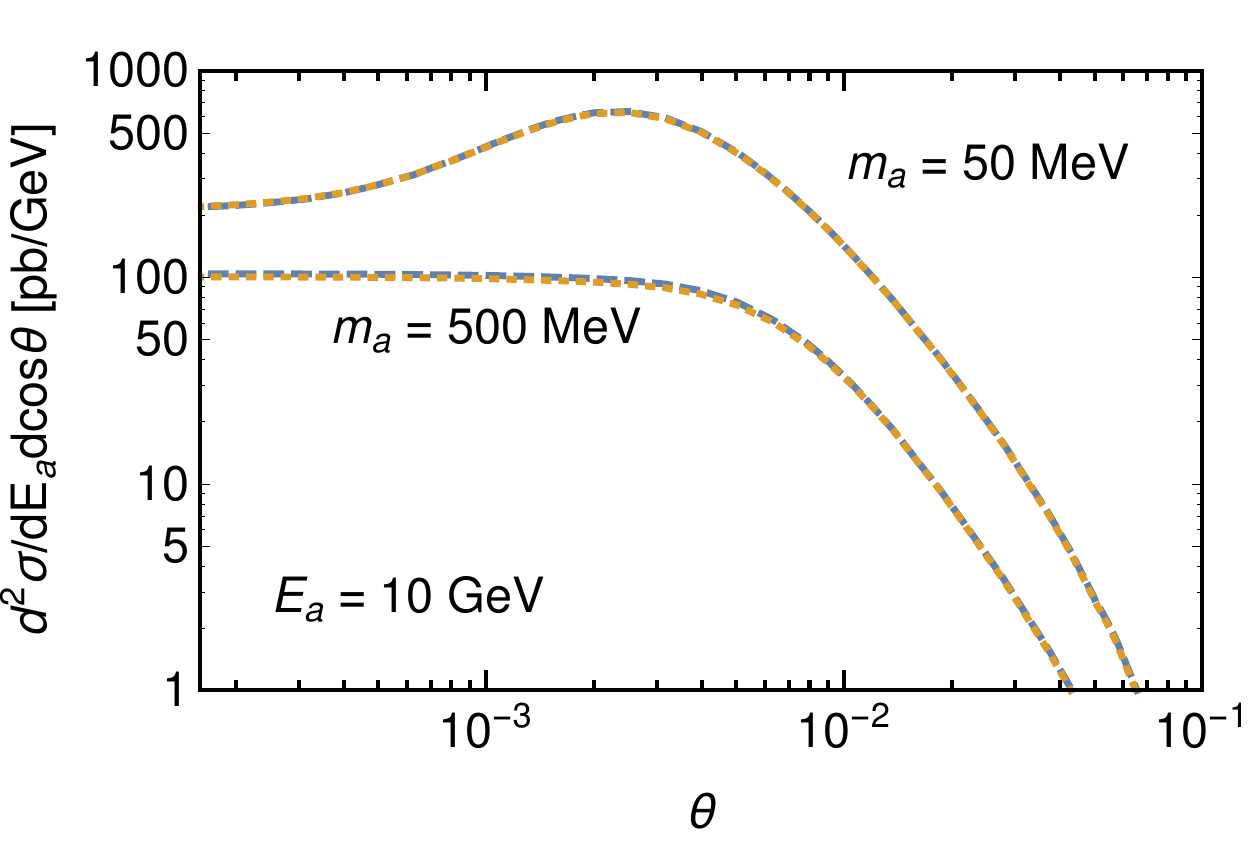}
\caption{Predictions for the differential ALP-production cross section in the laboratory frame as a function of $E_a$ (left) and $\theta$ (right) obtained in two different ways. Blue (dashed): Using equivalent photon spectra for both the proton beam and the target nuclei in the centre-of-mass frame and calculating the probability for photon fusion. Orange (dotted): Using equivalent photon spectra only for the proton beam in the laboratory frame and calculating the probability for ALP-emission and photon absorption. Note that, in contrast to figure~\ref{fig:alp-distribution}, we show $\mathrm{d}^2 \sigma_{pN} / \mathrm{d} E_a \, \mathrm{d} \cos \theta = (\sin \theta)^{-1} \, \mathrm{d}^2 \sigma_{pN} / \mathrm{d} E_a \, \mathrm{d} \theta$, which remains finite for $\theta \rightarrow 0$.}
\label{fig:agreement}
\end{figure}

The cross section for the process $p + N \rightarrow p + N + a$ is then given by
\begin{equation}
\frac{\mathrm{d}^2\sigma_{p N}}{\mathrm{d}E_a\mathrm{d}\cos{\theta}} = \frac{1}{2\pi\,E_\text{beam}} \int \mathrm{d}p_t^2 \, \mathrm{d}\phi \, \gamma_p(E_a/E_\text{beam}, q_t^2) \, \frac{\mathrm{d}\sigma_{\gamma N}}{\mathrm{d}\cos{\theta}} \; ,
\label{eq:method2}
\end{equation}
where we have made use of the fact that $\gamma_p(x, p_t^2) \mathrm{d}x = \frac{1}{E_\text{beam}} \, \gamma_p(E_\gamma/E_\text{beam}, p_t^2) \mathrm{d}E_\gamma$ and taken $E_\gamma \approx E_a$. Compared to eq.~(\ref{eq:method1}), the advantage of eq.~(\ref{eq:method2}) is that it only requires integration over one transverse momentum. However, in the latter case it is not possible to analytically perform the averaging over the angle of the photon in the transverse direction, so that both methods require a comparable computational effort. Moreover, as discussed in detail in appendix~\ref{app:crosssections}, eq.~(\ref{eq:method2}) is valid only for small angles and small $p_t$, whereas no such approximation was necessary for the derivation of eq.~(\ref{eq:method1}). Nevertheless, as long as these approximations are valid, the two methods should be in agreement. We confirm in figure~\ref{fig:agreement} that this is indeed the case.

To conclude this section, let us comment on the ALP angular distribution. First of all, we emphasise that in figure~\ref{fig:alp-distribution} we show $\mathrm{d}^2 \sigma_{pN} / \mathrm{d} E_a \, \mathrm{d} \theta = \sin \theta \, \mathrm{d}^2 \sigma_{pN} / \mathrm{d} E_a \, \mathrm{d} \cos \theta$, in order to reflect the larger solid angle at larger values of $\theta$. We find that (for a beam energy of $400\:\text{GeV}$) this distribution peaks at around $20\:\text{mrad}$, implying that an ideal detector should aim to cover this angle in order to capture the majority of the ALPs produced in the target. For comparison, we show $\mathrm{d}^2 \sigma_{pN} / \mathrm{d} E_a \, \mathrm{d} \cos \theta$ in figure~\ref{fig:agreement}, which remains finite for $\theta \rightarrow 0$. However, we observe that~--- at least for small ALP masses~--- even this distribution peaks at non-zero $\theta$. The reason is that, as can be seen from eq.~\eqref{eq:epa}, a photon with energy $\omega = x\,E_\text{beam}$ has a typical transverse momentum of order $q_t \sim x \, m_p$ and hence a typical angle of $q_t / \omega = m_p / E_\text{beam}$. For a beam energy of $400\:\text{GeV}$ the maximum is therefore found around $\theta \sim 2.5\,\text{mrad}$.\footnote{To produce an ALP with mass $m_A$ and energy $E_A$ requires a minimum momentum transfer of $q^2_\text{min} = m_A^4 / (4 \, E_A^2)$. If this quantity is comparable to $R_1^{-2}$ (where $R_1$ is the size of the target nucleus), any additional transverse momentum $q_t > 0$ will be strongly suppressed by the form factor of the target nucleus. As a result, the peak of the differential cross section for non-zero $\theta$ disappears for large ALP masses and small ALP energies.}

\section{Experimental event rates\label{sec:eventrates}}

Having calculated the rate and distribution of ALPs produced in the beam dump, we now proceed to calculating the probability that these ALPs lead to an observable signal in a given experiment. As the typical experimental set-up we assume that the target is immediately followed by an absorber of length $D$ and a decay volume of length $L$. At the far end of the decay volume there is a photon detector, which we take to be cylindrically symmetric with radius $R$, such that it covers an angle $\theta_\text{max} \approx R/(D + L)$ from the interaction point. A concrete example of such an experiment is NA62, which is shown in figure~\ref{fig:na62} and discussed in detail in section~\ref{sec:na62}.

The first ingredient for calculating the signal acceptance is the probability that the ALP passes through the absorber without decaying or interacting and then decays inside the decay volume. For an ALP with decay length $l_a$ this probability is given by
\begin{equation}
p(l_a) = \exp\left(-\frac{D}{l_a}\right) - \exp\left(-\frac{D+L}{l_a}\right) \; .
\end{equation}
For $L \gg l$ it is a good approximation to neglect the second term because almost all ALPs that reach the decay volume will decay before arriving at the detector. The simplest way to estimate the signal acceptance is then to assume that every ALP decaying in the decay region leads to an observable signal in the detector, provided the ALP angle satisfies $\theta < \theta_\text{max}$. We will use this approach in the following section for the analysis of CHARM and NuCal.

For present and future proton beam dump experiments, a more refined analysis strategy will be necessary in order to suppress potential backgrounds and allow the reconstruction of ALP properties in the case of a discovery. Most importantly, we require that both photons produced in the ALP decay are detected, i.e.\ we require two electromagnetic showers in the calorimeter coincident in time. The signal acceptance then needs to include
\begin{enumerate}
 \item the probability that both photons produced in the ALP decay reach the detector located at the far end of the decay volume;
 \item the probability that both photons are detected and give a signal that can be 
 distinguished from potential backgrounds.
\end{enumerate}

The first part depends mostly on the angle of the  ALP, $\theta$, and its boost $\gamma$,
which determines the typical opening angle between the two photons produced in the laboratory frame. 
However, there will also be some dependence on the ALP 
laboratory-frame decay 
length $l_a = \beta \, \gamma \, \tau$, because the separation between 
the two photons will be smaller the closer the ALP can get to the detector before decaying.
The second part of the probability can depend on the photon position and energy. More importantly 
it will depend on whether the separation between the two photons is large enough to identify 
two separate photons which are coincident in time.
Note that for a typical calorimeter the detection probability for
photons is very close to unity across the entire
energy range of interest. It is straight-forward to determine the detection probability as a function of 
$\theta, \gamma$ and $l_a$ from 
Monte-Carlo simulations of ALP decays in a given detector geometry. 
Nevertheless, before presenting the results of these simulations, it is useful to develop an intuitive understanding of the most important contributions.

First of all, if the direction of the ALP is such that $\theta > \theta_\text{max} \approx R/(D + L)$, 
it is impossible for both photons produced in the ALP decay to reach 
the detector, so that the detection probability vanishes: $p(\theta > \theta_\text{max}) = 0$. For $\theta < \theta_\text{max}$, on the other hand, the detection probability depends on the expected separation of the two photons. Since the decay of an ALP into photons is isotropic in the ALP rest frame, the distribution of the opening angle $\alpha$ between the two photons in the laboratory frame is given by
\begin{equation}
 \frac{\mathrm{d}N}{\mathrm{d}\alpha} = \frac{1}{4 \, \gamma \, \beta}\frac{\cos \alpha/2}{\sin^2 \alpha/2}\frac{1}{\sqrt{\gamma^2 \, \sin^2 \alpha/2 - 1}} \; ,
\end{equation}
which is strongly peaked towards the minimum opening angle $\alpha_\text{min} \simeq 2/\gamma$ (assuming $\gamma \gg 1$). If the ALP lifetime is short compared to the length of the decay volume, such that most ALPs will decay at the beginning of the decay volume, the typical separation of the two photons at the position of the detector will hence be $d_{\gamma\gamma} = \alpha_\text{min} \, L = 2 L / \gamma$.

This consideration enables us to constrain the acceptance as a function of the boost factor $\gamma$. 
Clearly, it is a necessary requirement
that $d_{\gamma\gamma} < d_\text{max} \equiv 2 \, R$. Consequently, we conclude $p(\gamma < L / R) = 0$. If $\theta$ is significantly different from zero, even larger boost factors will be required in order for 
both photons to reach the detector. 
On the other hand, we also require that the separation between the two photons is larger than a 
certain minimal distance $d_\text{min}$ to observe 
two individual photons. This requirement can be translated into an upper bound on the boost 
factor: $p(\gamma > 2 \, L / d_\text{min}) = 0$.

In summary, for short ALP decay length, $l_a \ll L$, we can roughly estimate the detection probability as
\begin{equation}
 p(l_a, \gamma, \theta) = \left\{
     \begin{array}{ll} \exp\left(-\frac{D}{l_a}\right) \; &  \text{for}\ \gamma > L / R, \ \gamma < 2 \, L / d_\text{min} \ \text{and} \ \theta < \theta_\text{max} \\
      0 & \text{otherwise}
     \end{array}
   \right.
\label{eq:acceptance}
\end{equation}
It is clear that this estimate is rather simplistic. The easiest way to obtain a more realistic estimate of the detection probability is to perform a Monte-Carlo simulation of ALP decays. The basic idea of such a simulation is to randomly determine the decay point and the direction of the emitted photons for a large number of ALPs with given properties $l_a$, $\gamma$ and $\theta$ and then determine the fraction of events where both photons hit the detector with the required separation.

Once the detection probability is known, we can calculate the number of predicted events by comparing the cross-section for ALP production with the 
cross-section for proton-nucleus scattering $\sigma_{pN}$. 
According to~\cite{Carvalho:2003pza}, this cross section is largely independent of the beam energy and can be parametrized as
\begin{equation}
 \sigma_{pN} = 53\:\text{mb} \times A^{0.77} \; ,
\end{equation}
where $A$ is the mass number of the nucleus. The number of ALP-induced events in a given experiment is then
\begin{equation}
 N = \frac{N_\text{pot}}{\sigma_{pN}} \int \frac{\mathrm{d}\sigma_f}{\mathrm{d}E_a \, \mathrm{d}\theta} \, \mathrm{d}E_a \, \mathrm{d}\theta \; ,
\label{eq:events}
\end{equation}
where $N_\text{pot}$ is the number of protons on the target material and
\begin{equation}
 \frac{\mathrm{d}\sigma_f}{\mathrm{d}E_a \, \mathrm{d}\theta} = p(l_a,\theta,\gamma) \cdot \frac{\mathrm{d}\sigma}{\mathrm{d}E_a \, \mathrm{d}\theta}
\label{eq:fiducial}
\end{equation}
is the fiducial cross section, which implicitly depends on the ALP mass and decay constant. Above, we have implicitly made the assumption that the target material is thick enough to absorb all protons to a good approximation.

Before we conclude this section, let us comment on the influence of the target material.
Since, according to the EPA, the photons probe the charge of the entire target nucleus, 
the ALP production cross-section is approximately proportional\footnote{In fact, the enhancement for heavy nuclei is somewhat suppressed due to their larger size. 
Smaller photon momenta are sufficient to resolve the substructure and there is thus an additional form 
factor suppression. Comparing copper and tungsten we find that the naive scaling law is shifted in favour 
of copper by about 10-20\%.}  to $Z^2$. According to eq.~(\ref{eq:events}), the ALP production cross section needs to 
be normalized to the total interaction rate of the protons with the target material. Thus, using the 
approximate expression for $\sigma_{pN}$ from~\cite{Carvalho:2003pza}, the predicted number of events for different target materials but equal acceptance and runtime can roughly be rescaled using a factor $Z^2/A^{0.77}$, which is equal to 30 for iron, 34 for copper, 52 for molybdenum, 99 for tungsten and 111 for lead. We note that for heavy ions one has an additional enhancement by the charge of the ion. However, typically the flux of ions is considerably lower at least partially eating up this advantage. Nevertheless it may be worthwhile to check for heavy ion experiments with an advantageous geometry.

\section{Existing constraints from past experiments\label{sec:constraints}}

In this section we derive constraints from past proton beam dump experiments using coherent production 
of ALPs via the Primakoff process and compare them with other exclusion bounds on the ALP parameter space from the literature. 
The two most constraining experiments are\footnote{We have also taken NOMAD into consideration
for which an analysis for ALPs from $\pi^0$ decays has been made previously \cite{Altegoer:1998qta}.
NOMAD has an on-axis detector located $835\:\text{m}$ away from a Beryllium target
and we estimate it to be less constraining than CHARM for the model we are considering.}
CHARM~\cite{Bergsma:1985qz}, which has recently been used to derive constraints on light scalars and pseudoscalars with couplings to quarks~\cite{Bezrukov:2009yw,Clarke:2013aya}, 
and NuCal~\cite{Blumlein:1990ay}, which has been reanalysed for the case of hidden photons~\cite{Blumlein:2011mv,Blumlein:2013cua}.

The CHARM experiment has performed a search for hidden particles decaying into photons based on a dataset of $N_\text{pot} = 2.4 \cdot 10^{18}$ protons on a copper target.
The detector was placed 480 metres away from the beam-dump and was 35 metres long. 
The detector was $3\:\text{m}\times3\:\text{m}$ in transverse dimension
and was placed $5\:\text{m}$ away from the beam axis, so that it covered about $10\%$ of {the full circle} around the beam axis. 
The CHARM experiment required a single electromagnetic shower in the detector and quotes a signal acceptance of $51\%$. 
The detection probability was therefore given by (see section~\ref{sec:eventrates})
\begin{equation}
 p(l_a, \theta)_\text{CHARM} = \left\{
     \begin{array}{ll} 0.05 \times \left[\exp\left(-\frac{480\:\text{m}}{l_a}\right) - \exp\left(-\frac{515\:\text{m}}{l_a}\right)\right] \ &  \text{ for } 0.0068 < \theta < 0.0126 \\
      0 & \text{ otherwise}
     \end{array}
   \right.
\end{equation}
Since CHARM observed 0~events, we set a bound at 90\% confidence level of $N_{\rm{det}}<2.3$~events~\cite{Clarke:2013aya}.

The NuCal experiment made use of the U70 proton beam with a beam energy of $70\:\text{GeV}$. The lower beam energy is compensated by the much smaller distance between target and detector of only $64\:\text{m}$ and the comparably large detector length of $23\:\text{m}$. We follow the analysis strategy of~\cite{Blumlein:2013cua}, requiring a minimum ALP energy of $E_a > 10\:\text{GeV}$. With this requirement, the detector acceptance is approximately constant and equal to $70\%$. Since the detector has a radius of $1.3\:\text{m}$, we then obtain
\begin{equation}
 p(l_a, \theta)_\text{NuCal} = \left\{
     \begin{array}{ll} 0.7 \times \left[\exp\left(-\frac{64\:\text{m}}{l_a}\right) - \exp\left(-\frac{87\:\text{m}}{l_a}\right)\right] \ &  \text{ for } \theta < 0.015 \\
      0 & \text{ otherwise}
     \end{array}
   \right.
\end{equation}
In a dataset of $N_\text{pot} = 1.7 \cdot 10^{18}$ protons on an iron target, NuCal observed 1 event compared to an expectation of 0.3 events. At 90\% confidence level, we can therefore exclude any parameter point predicting more than 3.6 events.

The parameter regions excluded by CHARM and NuCal are shown in figure~\ref{fig:exclusions}. We also show a compilation of other constraints on the ALP parameter space (taken from~\cite{Jaeckel:2015jla} and references therein), most notably the ones set by the electron beam dump searches SLAC 141~\cite{Riordan:1987aw} and SLAC 137~\cite{Bjorken:1988as}. As emphasised in section~\ref{sec:review} the sensitivity of the different experiments depends decisively on the overall geometry of the different set-ups and therefore varies significantly between the different experiments under consideration. 
As a result, our reanalysis of the constraints from CHARM and NuCal improves 
significantly upon existing constraints. Let us now discuss what further improvements can be achieved with present and near-future experiments.

\section{Projections for an ALP search at NA62}
\label{sec:na62}

\begin{figure}
\begin{picture}(300,100)(0,-20)
\thicklines
\put(45,-5){\line(1,0){130}}
\put(45,-5){\line(0,1){5}}
\put(45,-5){\line(0,-1){5}}
\put(175,-5){\line(0,1){5}}
\put(175,-5){\line(0,-1){5}}
\put(105,-15){$$D$$}
\put(175,-5){\line(1,0){222}}
\put(397,-5){\line(0,1){5}}
\put(397,-5){\line(0,-1){5}}
\put(265,-15){$$L$$}
 \includegraphics[width=1\textwidth]{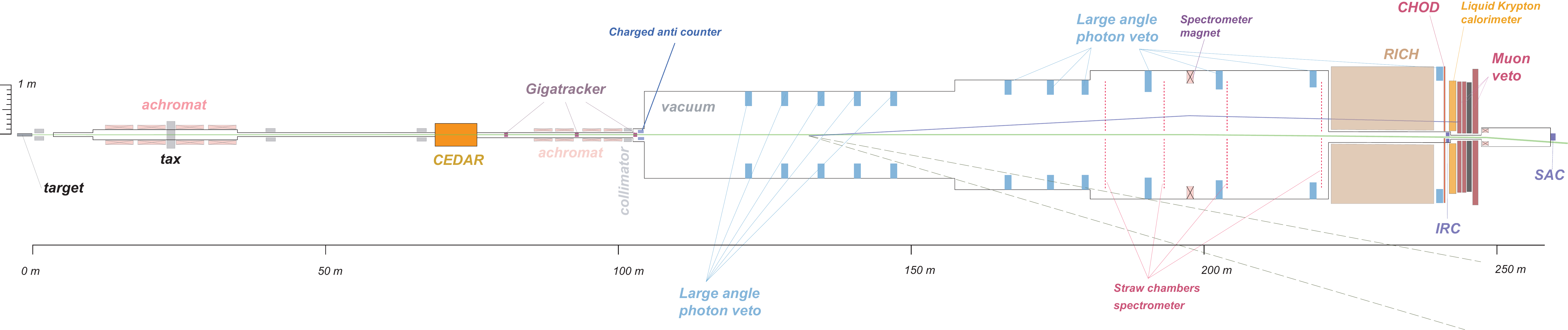}
\end{picture}
 \caption{Layout of NA62, sketch taken from~\cite{Valente:1293104}. 
 The SPS proton beam (from left), hits the target and the 
kaons of the secondary hadron beam are identified and measured
 before entering the vacuum decay region
 downstream. For a potential ALP search, 
 most preferentially the beam should directly impinge on the TAX (see text),
 the liquid krypton calorimeter for photon detection is placed approximately at a 
 distance of $241\:\text{m}$ behind the target. Overall exact geometric information is available at \cite{BEATCH}.
 The bar below the diagram indicates the effective length of the absorber $D$ and the decay volume $L$ we have used in our calculations.}
 \label{fig:na62}
\end{figure}
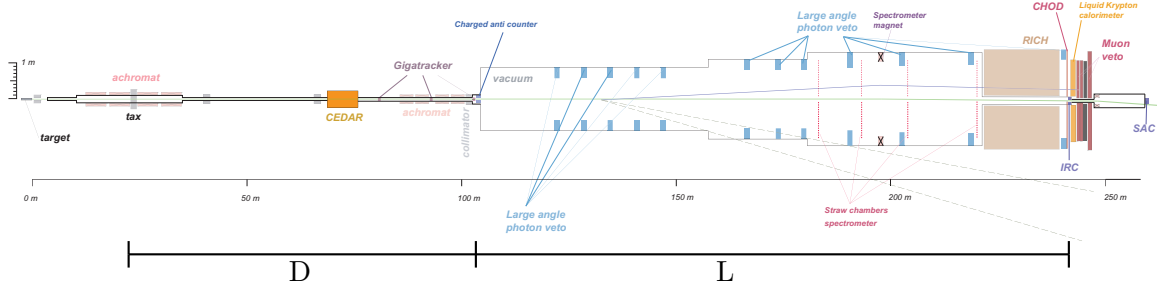

The fixed-target experiment NA62 at the CERN SPS aims to measure
the  $K^{+} \rightarrow \pi^{+} \nu \bar{\nu}$ branching ratio at the 10 \% level
within the next few years and is currently taking data. As $\pi^0$s
constitute a major background
for this measurement, overall hermetic detection of photons is crucial. Consequently,
NA62 could be adapted to search for ALPs decaying to two photons,
and we will elaborate on this possibility.

The NA62 experiment uses $400\:\text{GeV}$ protons from the SPS impinging on a beryllium target. The positively charged hadron beam of 75~GeV is then selected via an achromat: Only the wanted momentum component can pass through a set of apertures within
the so-called `Target Attenuator eXperimental areas' (TAXes)', see Chapter 2.1.2 of \cite{Hahn:1404985}. The particles
passing the apertures are then guided by magnetic elements to the decay volume downstream,
cf.\ figure~\ref{fig:na62} and \cite{Hahn:1404985} for details on the detector.

For a potential ALP search, two considerations for the primary beam are important
compared to `regular' data taking
\begin{enumerate}
\item To reach an acceptable background level, the experiment must be run in `dump mode'. 
Technically, this is possible by `closing' 
the two TAXes, which are 
approximately $25\:\text{m}$ behind the target. This means that the TAXes are positioned
such that no free aperture is being made available for main beam and all secondaries.
In this way, almost only muons and neutrinos are able to reach the decay volume (both TAXes are approximately $1.6\:\text{m}$ thick).
\item As argued in section~\ref{sec:eventrates}, a high-$Z$ fixed target is preferable
in producing ALPs through photon fusion. Removing the beryllium target
from the beam-path allows all 400GeV protons
to impinge directly onto the copper of the first TAX,
offset by $\sim 2\:\text{cm}$ downwards from the straight beam path due to the bend magnet.\footnote{In fact, 
both TAXes are equipped with two holed inserts (the apertures for the appropriate component of the
secondaries) and one $40\:\text{mm}$ diameter solid tungsten insert, respectively. Impinging on the 
full tungsten insert would be obviously be even more favourable than impinging on copper.
The solid tungsten insert would be large enough in diameter to absorb the full proton beam (which is 
of a couple of mm in transverse size). However the implemented range of movement for the TAXes does not allow to 
impinge on the tungsten of the first TAX directly. Doing so would require modifications in the target 
area and we do not consider this situation further. We also emphasise that, even in the configuration 
with the target in place, a sizable fraction ($\sim$40 \%) of the protons go through the target
without interacting and eventually impinge on the copper of the TAX.}
\end{enumerate}

After its production in the TAX, the ALP
needs to reach the decay volume (behind the `CHarged Anti Counter'). 
We therefore require that the ALP propagate along a distance
of $D=81$m before decay. As distance over which ALP decays can give a signal in the calorimeter 
we use $L=135$m.

As mentioned above, an excellent photon vetoing is crucial for the main physics goal of NA62.
For our estimate, we conservatively assume usage
of the Liquid Krypton Calorimeter (LKr) for photon detection only.
The LKr was previously used in NA48 and is described in detail in \cite{Fanti:2007vi}. 
A hole in the centre of the calorimeter hosts the vacuum tube for the beam ($8\:\text{cm}$ radius) and
the calorimeter itself is segmented in the transverse plane into cells of approximately $2\:\text{cm}$ by $2\:\text{cm}$.
Additional calorimeters (IRC, SAC) are installed to cover the calorimeter hole,
but as we will see below the presence of the LKr calorimeter hole has little influence on 
the ALP acceptance.

Let $\vec{r}_{1,2}$ be the two-vectors of the two photons on the calorimeter plane
and $r_{1,2}$ be their radial distance with respect to the beam axis.
Then, to summarize, the photons are required to be
\begin{enumerate}
\item within the calorimeter, thus $r_{1,2}< R = 113\:\text{cm}$;
\item far enough from the central hole to give a fully contained shower  $r_{1,2}> R_\text{min} = 15\:\text{cm}$;
\item separated by at least $d_\text{min} = |\vec{r}_{1}-\vec{r}_2|>10\:\text{cm}$ to avoid shower overlap.
\end{enumerate}
Note that the transverse shape of the LKr calorimeter is octagonal. 
For our estimates we implement the decay of ALPs in a toy Monte Carlo simulation unrelated 
to the NA62 software. In our Monte Carlo we approximate the LKr by a cylinder.

\begin{figure}[tb]
\centering
\includegraphics[width=0.47\textwidth]{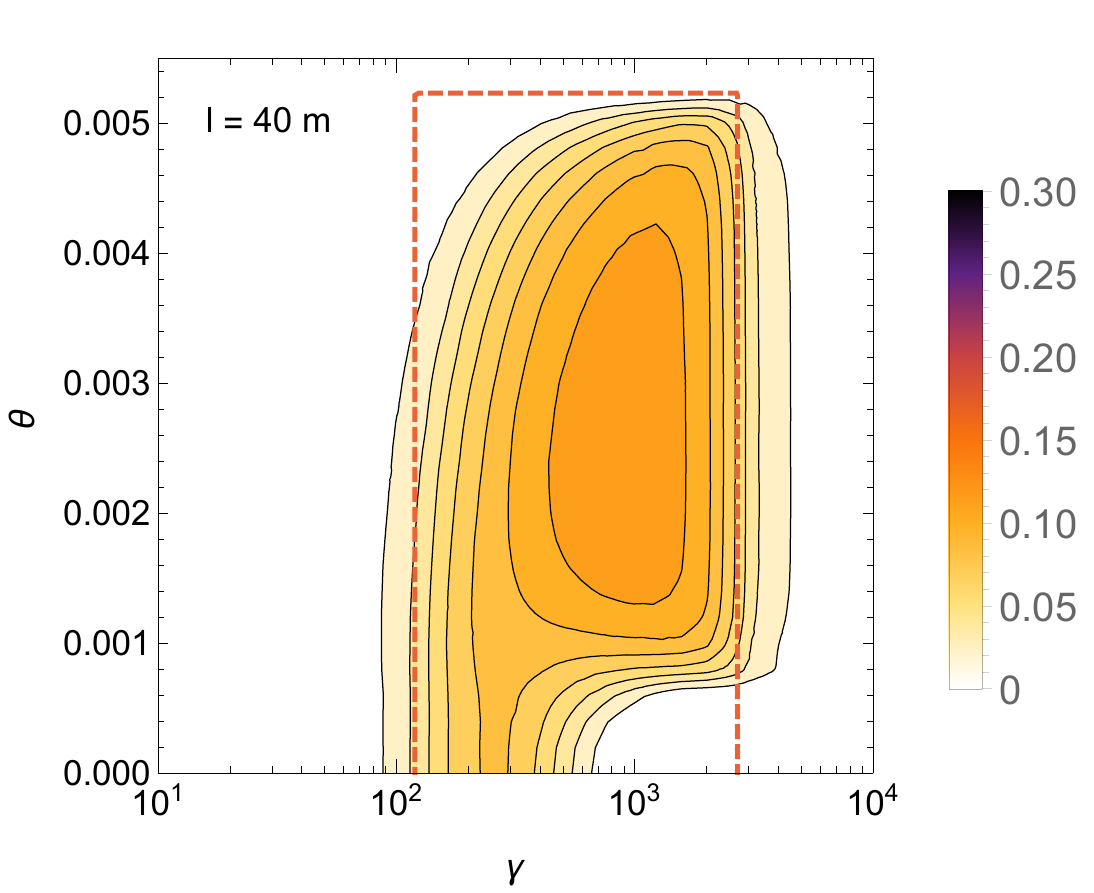}\hfill\includegraphics[width=0.47\textwidth]{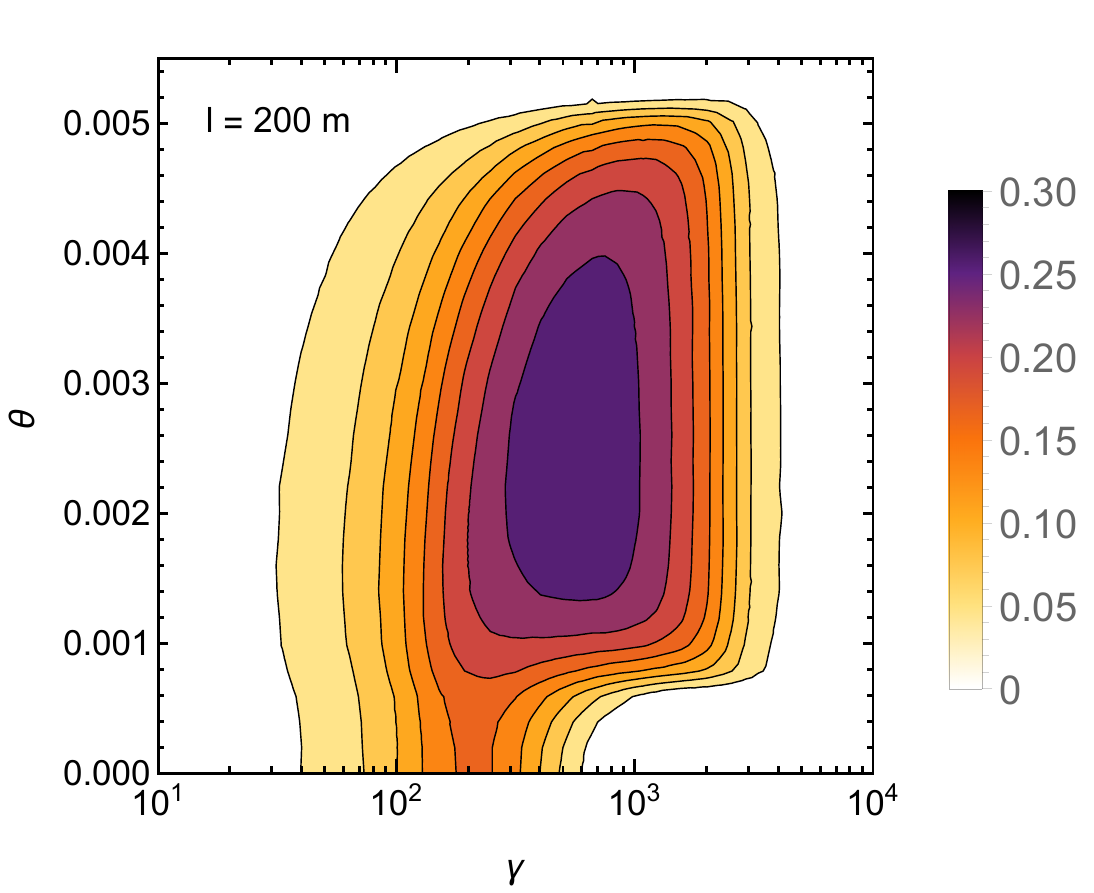}
\caption{Detection probability for ALPs in the NA62 detector as obtained from a Monte-Carlo simulation of ALP decays as a function of the ALP boost factor and angle for 
two different decay lengths $l_a$. In the left plot we also show the estimate of the acceptance window discussed in section~\ref{sec:eventrates} and summarised in eq.~(\ref{eq:acceptance}).}
\label{fig:acceptance}
\end{figure}

With the requirements above implemented, we can obtain a realistic estimate of the detection probability $p(l_a,\theta,\gamma)$
(an explicit dependence on $m_{a}$ and $\ga$ only arises after folding
the acceptance with the production cross-section). While it would be possible to consider photon losses in the spectrometer or Ring Imaging Cherenchov detector as well as detection efficiencies smaller than unity, we expect these effects to be small and therefore do not include them in our Monte-Carlo simulation. Figure~\ref{fig:acceptance} shows the resulting detection probability as a function of the ALP angle and its boost factor for different values of the decay length. 
The largest acceptances are found for off-angle ALPs and $l_a \sim D + L$, $\theta \sim 0.0025$ 
and $\gamma \sim 10^3$ and can reach up to $25\%$.

In the left panel of figure~\ref{fig:acceptance}, where the condition $l_a \ll L$ is satisfied, we also show for comparison the estimate for the acceptance region derived in section~\ref{fig:acceptance}. We find good agreement between our simple estimate and the results of the more detailed Monte-Carlo simulation. However, for longer decay lengths, as considered in the right panel, the acceptance region becomes much larger than predicted by our estimate, 
because a significant fraction of the ALPs decays closer to the detector. 
In addition to the features discussed in section~\ref{sec:eventrates} one can observe a loss of 
sensitivity for $\theta \approx 0$ and very large boost factors due to one or both photons 
being lost in the calorimeter hole. In agreement with expectations,
this effect becomes relevant if both $\theta$ and $\alpha_\text{min} / 2$ are small compared 
to $R_\text{min} / L$, implying $\theta < 0.0011$ and $\gamma > 900$. Since the ALP production
cross section is very small for $\theta \lesssim 10^{-3}$ (see figure~\ref{fig:alp-distribution}),
the presence of the calorimeter hole does not significantly affect the sensitivity of NA62 to ALPs.

\begin{figure}[tb]
\centering
\includegraphics[width=0.47\textwidth]{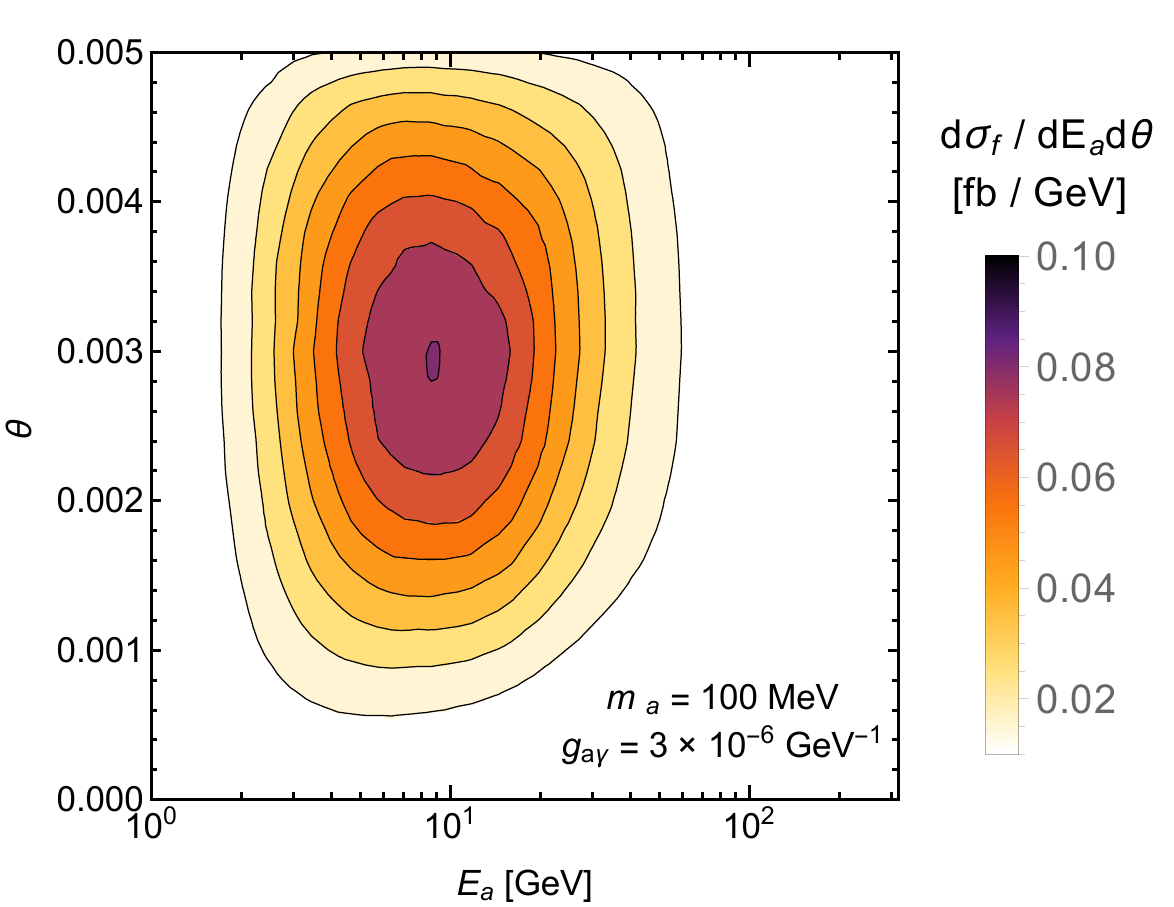}\hfill\includegraphics[width=0.47\textwidth]{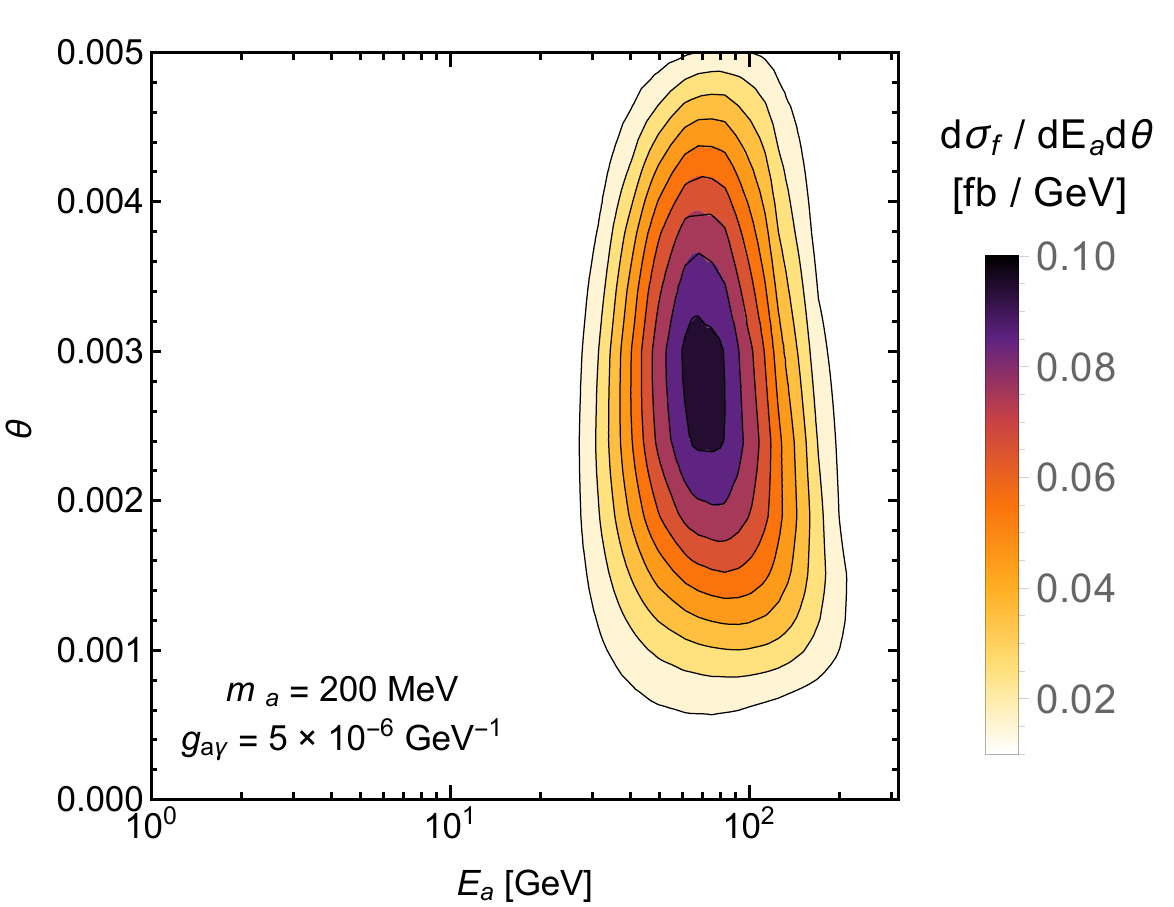}
\caption{Product of the differential ALP-production cross section (see figure~\ref{fig:alp-distribution}) and the detection probability in NA62 (see figure~\ref{fig:acceptance}) as a function of the ALP energy and angle for two different values of $m_{\rm a}$ and $\ga$, respectively.
}
\label{fig:crosssection}
\end{figure}

In fact, the angular distribution of ALPs produced via the Primakoff process peaks at somewhat larger angles and smaller boost factors than what can be detected by the NA62 detector, implying that an ideal detector should be somewhat larger and closer to the target. Nevertheless, high-energy ALPs are typically emitted with very small angle $\theta$, 
such that they match the acceptance window of the NA62 experiment. We show in figure~\ref{fig:crosssection} the fiducial cross section defined in eq.~(\ref{eq:fiducial}). While the left panel shows a typical distribution as obtained in large regions of parameter space, the right panel corresponds to ALPs with a rather short lifetime, such that only particles produced with very large boost factor reach the decay volume.

\begin{figure}[tb]
\centering
\includegraphics[width=0.47\textwidth]{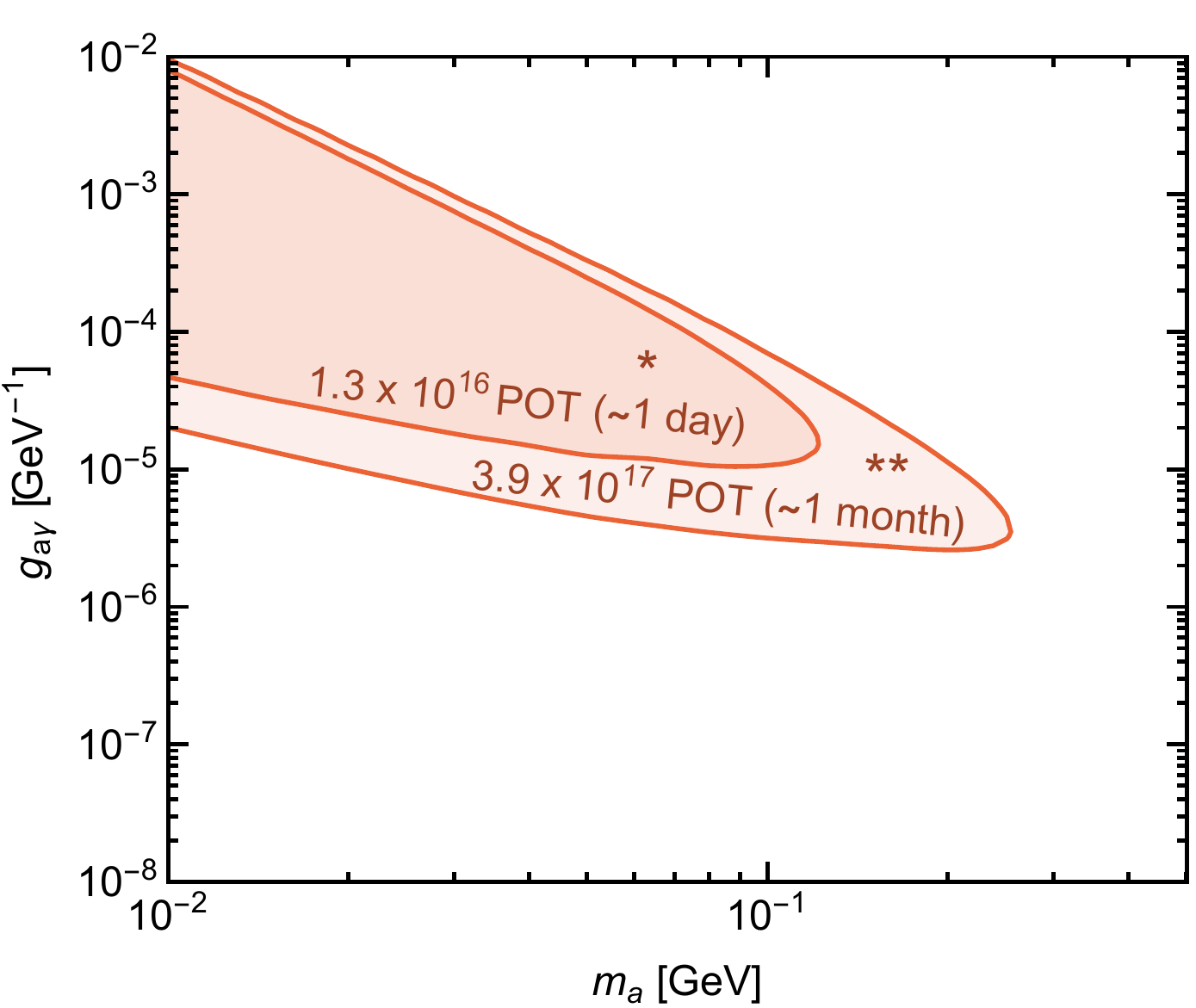}\hfill\includegraphics[width=0.47\textwidth]{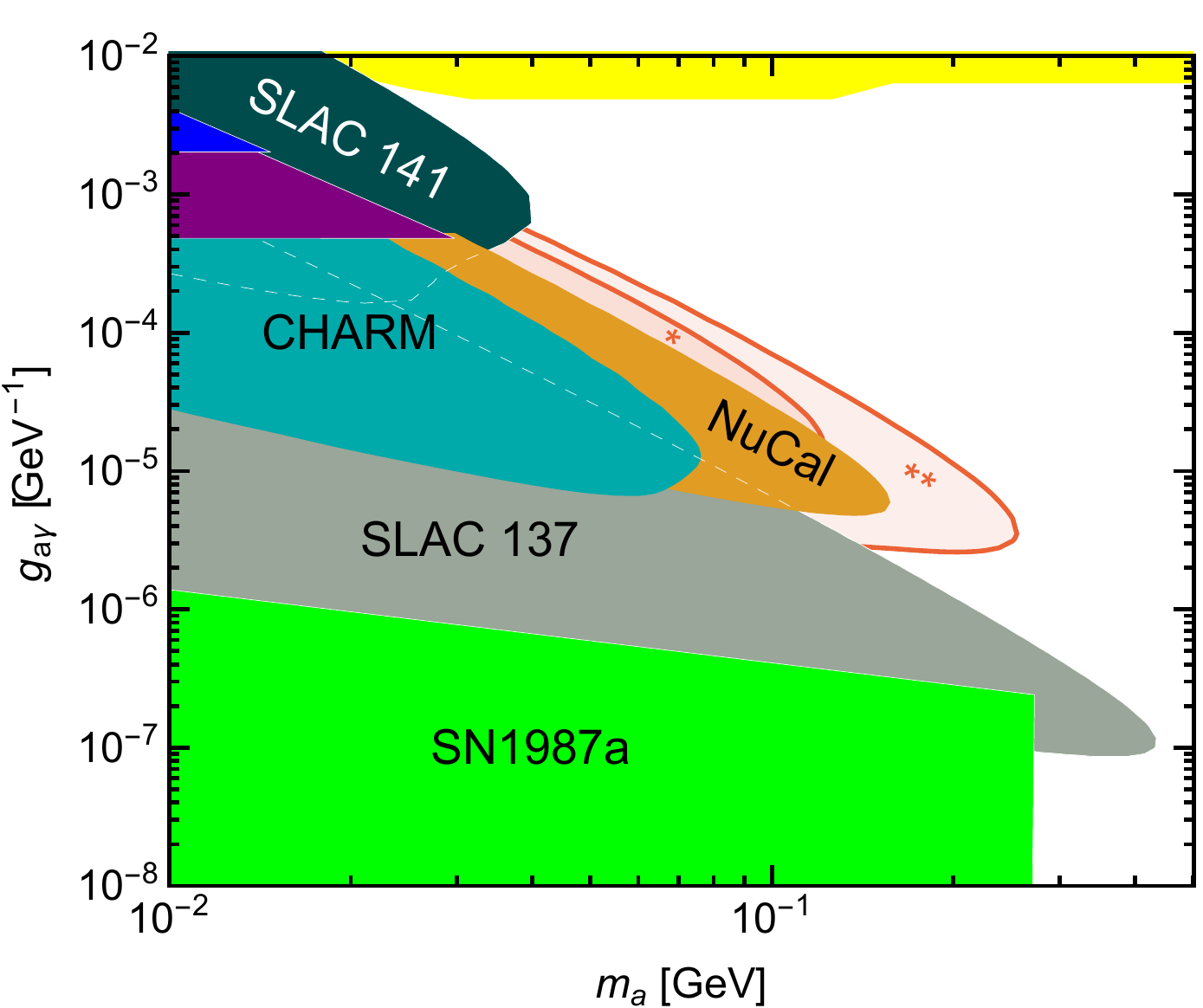}
\caption{
 Left: Projected sensitivity of NA62 in the ($m_a$, $\ga$) parameter plane for a number of 
 protons on target equivalent to one day (darker transparent red, marked with $\star$) and 
 equivalent to one month (lighter transparent red, marked with $\star \star$). 
 Right: Parameter regions of left-hand plot overlaid with previously performed experiments and 
 astrophysical constraints, see figure~\ref{fig:exclusions}.
}
\label{fig:projection}
\end{figure}

To calculate the predicted number of events in NA62, we take as nominal intensity $10^{12}$ protons per effective second on target. 
The effective flat top of the proton spill is approximately 3 seconds,                                                                                                                                                                                                                                                                                                                                                                                                                                                                                                                                                                                                           
2 times every super-cycle of 42 seconds, thus we have to multiply by an effective factor of 0.15. 
Within 24 hours, therefore $\sim 1.3 \times 10^{16}$ protons on target are available. 
Assuming that all backgrounds can be rejected, NA62 can probe all those points in 
the ($m_a$, $\ga$) parameter plane where at least three events are expected
from the production and decay of ALPs. Figure~\ref{fig:projection} shows the resulting expected sensitivity
for a data-taking period of 
1 day and 1 effective\footnote{Here, we do not account for detector- or beam-down-time.
So the actual run-time could be larger than a month.} month, respectively. As can be seen from the right 
panel, such a modest integrated intensity is fully sufficient to significantly improve upon existing 
upper limits from ALP-searches at beam dump 
experiments 
and probe presently unexplored regions of parameter space. We emphasize that doing so does not require any 
significant modification of the existing experimental setup.

\section{Sensitivity at the proposed SHiP facility \label{sec:SHiP}}

While NA62 offers a unique opportunity to probe unexplored parameter space with an existing experiment, 
there are various ways in which to further improve the sensitivity of future experiments. As discussed above,
the production rate is enhanced by a factor $Z^2 / A^{0.77}$ for heavy elements in the target. 
Furthermore, a shorter distance between the target and the beginning of the decay volume would allow to 
probe ALPs with shorter lifetimes, corresponding to larger couplings. Placing the detector closer to the target has the additional advantage that it is possible to cover somewhat larger ALP production angles and make use of the fact that the differential cross section typically peaks at around $\theta = 10\:\text{mrad}$.

All of these potential improvements have been incorporated in the SHiP proposal~\cite{Anelli:2015pba}, 
which aims to search for a variety of weakly-coupled low-mass states with very high 
intensity beams~\cite{Alekhin:2015byh}. In this section we present an estimate of the expected 
sensitivity of an ALP-search at SHiP.\footnote{Previous estimates of the sensitivity of SHiP for ALPs 
were based on incoherent production of ALPs, 
i.e.\ $q + \bar{q} \rightarrow a + \gamma$~\cite{Alekhin:2015byh}. 
Including the coherent production via Primakoff processes significantly increases the parameter
region that can be probed by SHiP.} We assume a total of $2 \cdot 10^{20}$ protons on a molybdenum target.

The decay volume is proposed to begin approximately at a distance of $70\:\text{m}$ from the target and 
is $50\:\text{m}$ long. 
We consider the detector to have a radius of $2.5\:\text{m}$, such that SHiP can cover production angles
up to $20\:\text{mrad}$. As for NA62, we require that both photons from the ALP decay reach
the detector 
and that the separation between them is at least $10\:\text{cm}$. Moreover, we require that the combined energy of both photons satisfies $E_\text{tot} > 3\:\text{GeV}$.

\begin{figure}[tb]
\centering
\includegraphics[width=0.5\textwidth]{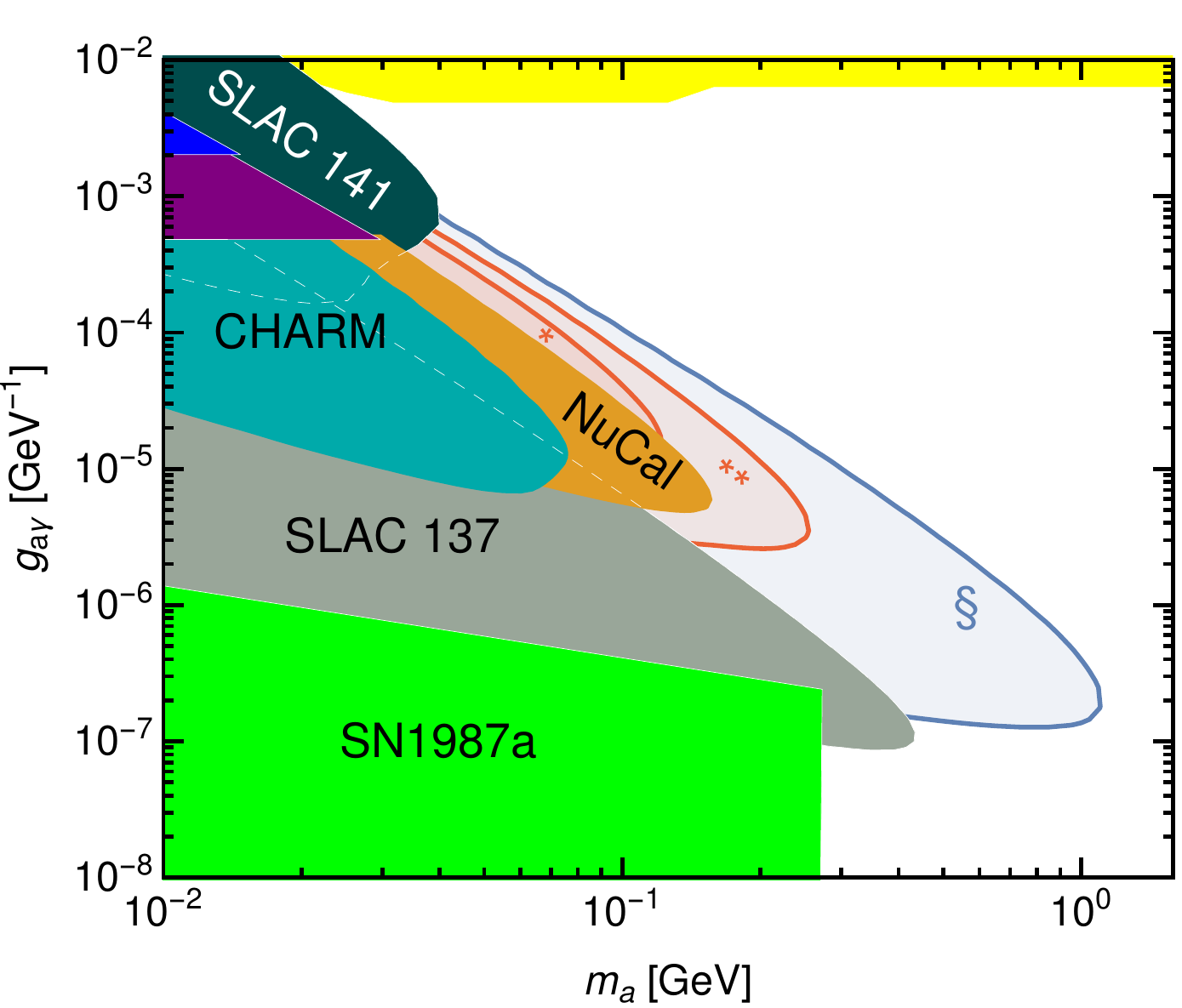}
\caption{
Projected sensitivity of SHiP, marked by \textsection, for $2 \cdot 10^{20}$ protons on target overlaid with figure~\ref{fig:projection}.
As in figure~\ref{fig:projection}, opaque regions correspond to existing limits,
transparent regions correspond to a proposed experimental reach based on assumptions as outlined in the text.
}
\label{fig:SHiP}
\end{figure}

An estimate of potential backgrounds is beyond the scope of this work. We therefore assume backgrounds
to be negligible and show the parameter region that would lead to at least 
three events in the SHiP experiment in figure~\ref{fig:SHiP}. The combination of an optimised design 
and a very large number of protons on target allows SHiP to explore large parameter regions inaccessible for other 
kinds of searches. The projected sensitivity extends all the way up to ALP masses of about $1\:\text{GeV}$ and ALP-photon couplings as small as $\ga = 10^{-7}\:\text{GeV}^{-1}$.

\section{Conclusions \label{sec:conclusions}}

Proton fixed target experiments provide exciting opportunities in
the search for new light and weakly coupled particles with masses in the MeV to GeV range. However, they confront us with the difficulty of colliding two composite particles, the proton and the even more complicated atoms of the target material. This raises the question whether one can provide a reliable calculation of the production rates. In this work we have addressed this problem by studying the Primakoff production of axion-like particles (ALPs) in proton fixed target experiments.

For a highly energetic proton beam of 400 GeV, as provided by the SPS, and ALP masses in the MeV to GeV mass range significant production is possible from photons with sufficiently high momentum to not be affected by the electron shell, and sufficiently low momentum that they do not resolve the proton and the nucleus. In this range a reliable calculation can be performed by taking into account simple electromagnetic form factors. As an additional benefit the scattering is then coherent over the whole nucleus and production cross sections are enhanced by a factor of the nuclear charge squared.

Importantly, we have provided predictions for the angular distribution of the ALP-production cross-section: Although transversal momenta of the produced ALPs are typically small, the detector in a beam dump experiment is placed far away from the target and therefore covers only a tiny angle from the production point. A precise determination of the expected spatial distribution for ALP-induced events and an accurate estimate of the geometric acceptance is therefore mandatory in preparing and analysing a real experimental run. Taking all these effects into account, we have shown that even with a rather modest beam-time requirement, the currently operating NA62 experiment would have a sizeable discovery potential for ALPs in the mass range of $\sim (30-200)$MeV. The proposed facility SHiP could extend this reach over the course of its running period up to masses of $1\:\text{GeV}$.

In the present work we have focussed on pseudoscalar ALPs that couple dominantly to photons. It is however straight-forward to generalise our results to scalar ALPs as well as ALPs with additional couplings to fermions as follows:
\begin{itemize}
 \item Writing the coupling between scalar ALPs and photons as $\frac{\ga}{2} \, a \, F^{\mu\nu} F_{\mu\nu}$, we obtain identical expressions for the ALP lifetime and the ALP production cross sections as for the case of the pseudoscalar. Our analysis therefore applies to this case as well.
 \item Even for ALPs with relatively large (derivative) couplings to fermions, the decay into photons will typically give the dominant contribution to the ALP decay length for $100 \: \text{MeV} \lesssim m_a \lesssim 2 \, m_\mu$~\cite{Dolan:2014ska}, which is the region of interest for NA62. For larger ALP masses, as potentially testable at SHiP, decays into muons (and, for scalar ALPs, mesons) can significantly reduce the ALP decay length and hence suppress event rates.
 \item In the presence of couplings to fermions, new production modes become available, such as ALP-strahlung, ALP-pion mixing or ALP production in flavour-changing rare decays. Depending on the details of the model under consideration these processes may significantly boost the ALP production rate and therefore enhance the sensitivity of beam dump experiments.
\end{itemize}

In conclusion, we have worked out a detailed example how physics beyond the SM at comparably low mass could be probed with existing set-ups with comparably low effort, providing a complementary window to the opportunities that the highest-energy accelerators are offering us.

\acknowledgments

We would like to thank Walter Bonivento, David d'Enterria, Javier Redondo and Pedro Schwaller for useful
discussions, and Johannes Bl\"umlein, Matthew Dolan and Tommaso Spadaro for insightful comments on the manuscript. BD acknowledges very helpful conversations within the NA62 collaboration, particularly with Brigitte Bloch-Devaux, Augusto Ceccucci, Luigi Di Lella, Niels Doble, Lau Gatignon, Evgueni Goudzovski, Matthew Moulson, Mathieu Perrin-Terrin, Mauro Raggi, Giuseppe Ruggiero, Tommaso Spadaro and Paolo Valente. JJ would like to thank the IPPP for a very enjoyable visit. FK and KSH are grateful to the CERN Theory Division for hospitality. This work is supported by the German Science Foundation (DFG) under the Collaborative Research Center (SFB) 676 ``Particles, Strings and the Early Universe'', the Transregio TR33 ``The Dark Universe'', as well as the ERC Starting Grant `NewAve' (638528).

\appendix

\section{Cross sections}
\label{app:crosssections}

Let us first calculate the cross section for the $2\rightarrow1$ process $\gamma + \gamma \rightarrow a$. The matrix element for the photon-photon-ALP vertex is given by
\begin{equation}
 \mathcal{M} = \ga \epsilon^{\mu\nu\rho\sigma} \epsilon_\mu(q_1) \epsilon_\nu (q_2) q_{1,\rho} q_{2,\sigma}
\end{equation}
where $\epsilon(q)$ is the photon polarisation vector. Averaging $|\mathcal{M}|^2$ over initial state polarisations gives
\begin{align}
 \frac{1}{4} \sum |\mathcal{M}|^2 & = \frac{\ga^2}{4} \epsilon^{\mu\nu\rho\sigma} \epsilon_{\mu\nu\kappa\lambda} q_{1,\rho} q_{2,\sigma} q^{1,\kappa} q^{2,\lambda} \nonumber \\
 & = \frac{\ga^2}{2} ((q_1 \cdot q_2)^2 - q_1^2 q_2^2) \nonumber \\
 & = \frac{\ga^2}{8} \sqrt{s}^4 \; .
\end{align}
We hence obtain
\begin{align}
 \sigma(\gamma \gamma \rightarrow a) = & \frac{1}{2 E_1 2 E_2 |v_1 - v_2|} \nonumber \\ & \times \int \frac{\mathrm{d}^3 k}{(2\pi)^3} \frac{1}{2 E_a} |\mathcal{M}(q_1,q_2 \rightarrow k)|^2 (2\pi)^4 \delta^{(4)}(q_1+q_2-k) \nonumber \\
= & \frac{\pi \ga^2 \sqrt{s}^4}{64 \, E_1 \, E_2 \, (E_1 + E_2)}\delta(E_1 + E_2 - E_a) \; .
\end{align}
Evaluating this expression in the centre-of-mass frame yields
\begin{equation}
 \sigma(\gamma \gamma \rightarrow a) = \frac{\pi\,\ga^2\,m_a}{16} \delta(\sqrt{s} - m_a) \; .
\end{equation}

\bigskip

We now want to calculate the cross section for the $2 \rightarrow 2$ process $\gamma + N \rightarrow a + N$, allowing non-vanishing transverse momentum for both the incoming photon and the outgoing ALP. In the lab frame (where the incoming nucleus is at rest), we adopt the following choice of coordinates:
\begin{equation}
p_\gamma = \begin{pmatrix} E_\gamma \\ p_t \sin \phi \\ p_t \cos \phi \\ p_z \end{pmatrix} \, , \qquad p_a = \begin{pmatrix} E_a \\ 0 \\ k_a \cos \theta \\ k_a \sin \theta \end{pmatrix} \, .
\end{equation}
The 4-momentum of the outgoing nucleus is then fully determined by energy and momentum conservation. Note also, that we require $E_\gamma^2 = p_t^2 + p_z^2$ and $E_a^2 = k_a^2 + m_a^2$, i.e.\ we require all particles to be on-shell.\footnote{In principle, the incoming photon can be off-shell. However, the dominant contribution will result from incoming photons that are (almost) on-shell, so the chosen approximation should be valid.} The photon momentum $E_\gamma$ is then fully determined by specifying the energy and direction of the outgoing ALP, the transverse momentum $p_z$ of the photon and the angle $\phi$ between the transverse momentum of the photon and the transverse momentum of the ALP.

Under the assumption that $\theta \ll 1$ and $m_a, \, p_t \ll E_a, \, m_N$, we obtain to order\footnote{Note that the term proportional to $m_a^4$ is typically tiny compared to the terms proportional to $p_t^2$ and $\theta^2$. We keep these terms nevertheless in order to be able to compare our final result with the one obtained in~\cite{Bjorken:1988as} for $p_t = 0$.} $m_a^4$, $p_t^2$ and $\theta^2$
\begin{equation}
 E_\gamma = E_a + \frac{m_a^4}{8 \, E_a^2 \, m_N} + \frac{p_t^2}{2 \, m_N} + \frac{E_a^2 \theta^2}{2 m_N} - \frac{ E_a \, p_t \, \theta \, \cos{\phi}}{m_N} \;.
\end{equation}
It is then possible to calculate all products of four-vectors at this order. In particular, we find
\begin{equation}
 t = (p_\gamma - p_a)^2 = -\frac{m_a^4}{4 \, E_a^2} - p_t^2 + 2 \, E_a \, p_t \, \theta \, \cos \phi - E_a^2 \, \theta^2 \; .
\end{equation}
Finally, we obtain for the squared matrix element
\begin{equation}
\sum_\text{spins} |\mathcal{M}|^2 = \frac{2 \pi \, \alpha \, \ga^2 \, m_N^2 \, (- 4 \, E_a^2 \, t - m_a^4)}{t^2} \; . 
\end{equation}
To obtain the scattering cross section, we need to multiply this expression with the nuclear form factor $Z^2 \, F(|t|)^2$ and a phase space factor
\begin{align}
\frac{\mathrm{d}\sigma_{\gamma N}}{\mathrm{d}\cos{\theta}} & = \sum_\text{spins} |\mathcal{M}|^2 \, Z^2 \, F(|t|)^2 \, \frac{1}{32 \pi \, m_N^2} \nonumber \\
& = \frac{\alpha \, \ga^2 \, (- 4 \, E_a^2 \, t - m_a^4)}{16 \, t^2} F(|t|)^2\; .
\end{align}
Reassuringly, this expression reduces to the one from~\cite{Bjorken:1988as} upon setting $p_t = 0$.

\section{Angular averaging}
\label{app:angular}

Let us consider two random 2-dimensional vectors $\mathbf{u}$ and $\mathbf{v}$. The probability distribution of the magnitude of these two vectors is given by the functions $f_1(u^2)$ and $f_2(v^2)$. The directions $\phi_u$ and $\phi_v$ of the two vectors are evenly distributed (i.e.\ isotropic). We are now interested in determining the probability distribution for the magnitude of $\mathbf{w} = \mathbf{u} + \mathbf{v}$. As a first step we define
\begin{equation}
 \tilde{f}_u (u^2, \phi_u) = \frac{1}{2\pi} f_u(u^2), \quad \tilde{f}_v (v^2, \phi_v) = \frac{1}{2\pi} f_v(v^2) \; .
\end{equation}
The distribution $f(w^2)$ can now be written as
\begin{align}
 f(w^2) & = \int \mathrm{d}u^2 \mathrm{d}\phi_u \, \tilde{f_u}(u^2, \phi_u) \int \mathrm{d}v^2 \mathrm{d}\phi_v \, \tilde{f_v}(v^2, \phi_v) \delta\left(w^2 - |\mathbf{u} + \mathbf{v}|^2\right) \nonumber \\
 & = \frac{1}{4\pi^2} \int \mathrm{d}u^2 \mathrm{d}\phi_u \, f_u(u^2) \int \mathrm{d}v^2 \mathrm{d}\phi_v \, f_v(v^2) \delta\left(w^2 - |\mathbf{u} + \mathbf{v}|^2\right) 
\; . 
\end{align}
Now we define the new angular variables $\Delta \phi = \phi_v - \phi_u$ and $\bar{\phi} = (\phi_v + \phi_u)/2$, noting that $\mathrm{d}\Delta \phi \mathrm{d} \bar{\phi} = \mathrm{d}\phi_u \mathrm{d}\phi_v$. Clearly, $|\mathbf{u} + \mathbf{v}|^2$ is independent of $\bar{\phi}$ and hence we can directly perform the integration:
\begin{equation}
 f(w^2) = \frac{1}{2\pi} \int \mathrm{d}u^2 \, f_u(u^2) \int \mathrm{d}v^2 \, f_v(v^2) \int \mathrm{d}\Delta\phi\ \, \delta\left(w^2 - |\mathbf{u} + \mathbf{v}|^2\right) 
\; . 
\end{equation}
The integral over $\Delta \phi$ vanishes unless $u$, $v$ and $w$ satisfy the triangle inequalities: $|u - v| \leq w \leq u + v$. With the range of integration thus restricted, we can solve
\begin{align}
 w^2 & = u^2 + v^2 + 2 u v \cos \Delta \phi \nonumber \\
\Rightarrow \cos \Delta \phi & = \frac{w^2 - u^2 - v^2}{2 u v} \nonumber \\
\Rightarrow \sin \Delta \phi & = \frac{\sqrt{2 w^2(u^2 + v^2) - w^4 - (u^2-v^2)^2}}{2 u v} \; .
\end{align}
Now we need to differentiate the argument of the $\delta$-function with respect to $\Delta \phi$ and substitute this solution:
\begin{align}
 \frac{d}{d\Delta\phi} \left(w^2 - u^2 + v^2 + 2 u v \cos \Delta \phi\right) & = 2 u v \sin \Delta \phi \nonumber \\
 & = \sqrt{2 w^2(u^2 + v^2) - w^4 - (u^2-v^2)^2}
\end{align}
Finally, we note that the argument of the $\delta$-function has two roots within $\left[0,2\pi\right]$, leading to an additional factor of 2. We thus obtain
\begin{equation}
 f(w^2) = \frac{1}{\pi} \int_{|u - v| \leq w \leq u + v} \mathrm{d}u^2 \mathrm{d}v^2 \frac{f_u(u^2)\, f_v(v^2)}{\sqrt{2 w^2(u^2 + v^2) - w^4 - (u^2-v^2)^2}}
\; . 
\end{equation}
It is an easy but useful cross check to integrate the right-hand side over $w^2$ to make sure that $f(w)$ is correctly normalised.

In some applications it may be useful to write the range of integration in a slightly different way. Clearly, we can integrate separately over the two regions $u < v$ and $u > v$. In the first case, the triangle inequalities imply $v^2 \geq w^2/4$ and $u^2 \geq w^2 + v^2 - 2 v w$. The integration can hence be written as
\begin{align}
 f(w^2) = & \quad \frac{1}{\pi} \int_{w^2/4}^\infty \mathrm{d}v^2 \int_{w^2 + v^2 - 2 v w}^{v^2} \mathrm{d}u^2 \frac{f_u(u^2)\, f_v(v^2)}{\sqrt{2 w^2(u^2 + v^2) - w^4 - (u^2-v^2)^2}} \nonumber \\
 & + \frac{1}{\pi} \int_{w^2/4}^\infty \mathrm{d}u^2 \int_{w^2 + u^2 - 2 u w}^{u^2} \mathrm{d}v^2  \frac{f_u(u^2)\, f_v(v^2)}{\sqrt{2 w^2(u^2 + v^2) - w^4 - (u^2-v^2)^2}}
\; . 
\end{align}

\bibliography{ALPtraum_colored}

\end{document}